\documentclass[%
 reprint,
 superscriptaddress,
nofootinbib,
 amsmath,amssymb,
 aps,
floatfix,
]{revtex4-2}

\usepackage{physics}
\usepackage{graphicx}
\usepackage{dcolumn}
\usepackage{bm}
\usepackage{hyperref}

\usepackage{braket}
\usepackage{xstring}
\usepackage{xcolor}
\usepackage[normalem]{ulem}

\graphicspath{{./figs/}}
\newcommand\GM[1][]{{\rm GM}^{\IfInteger{#1}{(#1)}{\mathtt{(#1)}}}}
\renewcommand\S[1][]{S^{\IfInteger{#1}{(#1)}{\mathtt{(#1)}}}}
\newcommand\Z[1][]{Z^{\IfInteger{#1}{(#1)}{\mathtt{(#1)}}}}

\begin{document}

\title{Symmetry-resolved genuine multi-entropy: \\ Haar random and graph states}

\author{Norihiro Iizuka}
\email{iizuka@phys.nthu.edu.tw}
\affiliation{Department of Physics, National Tsing Hua University, Hsinchu 300044, Taiwan}
\affiliation{Yukawa Institute for Theoretical Physics, Kyoto University, Kyoto 606-8502, Japan}

\author{Simon Lin}
\email{simonlin@nyu.edu}
\affiliation{New York University Abu Dhabi, P.O. Box 129188, Abu Dhabi, United Arab Emirates}

\date{\today}

\begin{abstract}
We study the symmetry-resolved genuine multi-entropy, a measure that captures genuine multi-partite entanglement, in Haar random states and random graph states in the presence of a conserved quantity. For Haar random states, we derive explicit formulae for the genuine multi-entropy under a global $U(1)$ symmetry in the thermodynamic limit, and find that its dependence on subsystem sizes closely resembles that of fully Haar random states without a conserved charge. We also perform numerical analyses, focusing on spin systems for both Haar random and graph states. For random graph states, our numerical analyses reveal distinctive features of their multi-partite entanglement structure and we contrast them with the Haar random case.  
\end{abstract}

\maketitle


\section{Introduction}
\label{sec:intro}

Entanglement is the defining feature of a quantum system.
Suppose that we have a system $M$ 
which we further divide into two subsystems $M=A\cup B$. We can quantify the amount of entanglement between $A$ and $B$ by calculating its \emph{entanglement entropy}:
\begin{equation}
  S(A) = -\tr (\rho_A\log \rho_A),
\end{equation}
where $\rho_A = \tr_B \rho_{AB}$ is the reduced density matrix of system $A$.
The entanglement entropy is a ubiquitous diagnostic: it obeys area laws \cite{Bombelli:1986rw, Srednicki:1993im, Callan:1994py} and distinguishes quantum phases via entanglement spectra \cite{Li:2008kda}, it is geometrized by Ryu–Takayanagi in holography \cite{Ryu:2006bv,Ryu:2006ef}, and for Haar random states it yields the universal Page curve \cite{Page:1993df, Page:1993wv}.

However, the entanglement entropy is a rather coarse measure for the entanglement structure of the system.
In particular, it is only sensitive to the \emph{bipartite entanglement} between system $A$ and $B$ in a pure state, while being agnostic to the much richer multi-partite entanglement structure of the system.
A typical quantum state in a generic system contains more than bipartite entanglement, and measures built solely from entanglement entropies can fail to detect these signatures of multi-partite entanglement. A nice example to illustrate this is the following two states \cite{PhysRevLett.86.5184, Gadde:2022cqi}: 
\begin{align}
\ket{\psi_1} &= \frac{1}{\sqrt{3}}\left( \ket{\uparrow \uparrow \downarrow} + \ket{\uparrow \downarrow \uparrow} + \ket{\downarrow \uparrow \uparrow}\right) \\
\ket{\psi_2} &= \frac{1}{\sqrt{3}}\left( \ket{\uparrow \uparrow \uparrow} + \sqrt{2} \ket{\downarrow \downarrow \downarrow}\right) 
\end{align}
Although the reduced density matrices of these two states are different, they share the same spectrum, {\it i.e.}, the same set of eigenvalues.
Consequently, any entanglement entropy or R\'enyi measure, which depends solely on the eigenvalue spectrum, cannot distinguish between the two states.
However, higher-partite quantities that are more sensitive than spectral data, such as the multi-entropy or the entanglement negativity, can differentiate between  them. 

Multi-entropies, or more generally, R\'enyi multi-entropies are generalizations of the R\'enyi  entanglement entropy into higher $\mathtt{k}$-partite\footnote{In the literature, such subdivisions are often referred to as $\mathtt{q}$-partite, however in this letter we use $\mathtt{k}$-partite instead, reserving $q$ for charge-related quantities.} ($\mathtt{k}>2$) subdivisions \cite{Gadde:2022cqi,Gadde:2023zzj}. They are constructed from symmetric contractions of the density matrix and is sensitive to higher-partite entanglements in the system. However, the multi-entropy does not always capture truly or genuine higher-partite entanglement by itself, since it is also sensitive to lower-partite correlations.
For example, the $\mathtt{k} = 3$ multi-entropy can be nonzero even for a state that contains only bipartite entanglement (such as an Bell-pair state).
Thus, one cannot determine whether a nonzero value of the $\mathtt{k} > 2$ multi-entropy arises from bipartite correlations or from genuinely higher $\mathtt{k} > 2$ multi-partite entanglement.
To resolve this issue, one can define the $k$-partite \emph{genuine} multi-entropies ($\GM[k]$) \cite{Iizuka:2025ioc,Iizuka:2025caq}\footnote{For the special case $\mathtt{k}=3$, it was already pointed out that a genuine three-party combination can be identified; see \cite{Penington:2022dhr,Gadde:2023zzj,Harper:2024ker,Liu:2024ulq}, in particular the quantity $\kappa$ discussed in \cite{Harper:2024ker}. To the best of our knowledge, however, a systematic formulation of genuine multi-entropy for general $\mathtt{k}$ was first developed in Refs.~\cite{Iizuka:2025ioc,Iizuka:2025caq}.}
, by systematically subtracting all the contributions from lower-partite entanglements such that the genuine multi-entropies are only sensitive to $k$-partite entanglements. These new measures have recently been used to argue the necessity of higher-partite entanglement in holographic systems \cite{Iizuka:2025bcc, Akers:2019gcv,Akers:2024pgq,Hayden:2021gno,Li:2025nxv}. For recent developments of GM, see \cite{Harper:2025uui,Iizuka:2025elr,Berthiere:2025toi,Balasubramanian:2025hxg,Bao:2025psl, Akella:2025owv,Yuan:2025dgx}. 

When the system of interest possesses a conserved charge $Q$, its Hilbert space decomposes into distinct sectors labeled by the total charges of the system. This is often the case in real-world scenarios, in which the charge $Q$ can be the total energy or the electric charge of the system.
It is natural to ask how does the entanglement structure change when one restricts to such a charged sector, or in other words, under \emph{symmetry-resolution}. The study of symmetry-resolved entanglement has been under active research in the recent years \cite{Goldstein:2017bua,Xavier:2018kqb,Bonsignori:2019naz,Fraenkel:2019ykl,Murciano:2019wdl,Feldman:2019upn,Murciano:2020vgh,Capizzi:2020jed,Lau:2022hvc,Neven:2021igr,Ares:2022koq}, especially for the study of the Page curve  \cite{Bianchi:2019stn,Murciano:2022lsw}.
However, most literature on this line of research has been focusing on bipartite entanglement measures\footnote{See Ref.~\cite{Berthiere:2023gkx} for results on two symmetry-resolved tripartite entanglement measures: computable cross-norm (CCNR) negativity and reflected entropy. Also see Ref.~\cite{Jain:2025aqs} on the symmetric resolution of the Q-measure of global entanglement \cite{Meyer:2001ckx}.}, and a systematic treatment for multi-partite entanglement is lacking.

In this letter, we initiate a first step in the systematic study of symmetry-resolved multi-partite entanglement by investigating the behavior of GM under symmetry resolution. 
We employ spin chain systems endowed with a global $U(1)$ symmetry as tractable toy models to enable explicit calculations of the symmetry-resolved genuine multi-entropy.
Our broader motivation lies in black hole physics, especially in understanding black hole (genuine) multi-entropy curves \cite{Iizuka:2024pzm}, where conservation laws such as total electric charge and energy are intrinsic.
In such contexts, symmetry resolution provides a natural framework to isolate how conservation constraints modulate genuine multi-partite entanglement.

It would be interesting to investigate how the (genuine) multi-entropy curves are modified 
(1) in the presence of a conserved quantity, and 
(2) when the ensemble is restricted to special classes of states. 
In this letter, 
following the approaches of \cite{Bianchi:2019stn,Murciano:2022lsw}, we provide analytical formulae of symmetry-resolved multi-entropy for \emph{Haar random states} (states described by a random vector in the Hilbert space) in the thermodynamic limit, which we compare with numerical evaluations. We also look at a different class of states known as \emph{graph states}. They are stabilizer states built from acting with commuting Clifford operators according to a graph. 

This letter is structured as follows: In Sec.~\ref{sec:review} we briefly review the notion of GME and the symmetry resolution of entanglement entropy. In Sec.~\ref{sec:SRGM} we study symmetry-resolved genuine multi-entropy curves for the ensemble of Haar random states and random graph states. We conclude with several remarks in Sec.~\ref{sec:summary}. Some technical derivations are relegated to the appendix.

\section{Prerequisites}
\label{sec:review}

\subsection{Symmetry-resolved entanglement entropy}
We begin by reviewing symmetry resolution of bipartite entanglement entropy for Haar random states \cite{Bianchi:2019stn,Murciano:2022lsw}.
To make the discussion concrete, we will throughout consider a system of $N$ spins with a global $U(1)$ charge $Q$, defined as the expectation value of the following operator:
\begin{equation}
  \hat{Q} = \frac{1}{2}\sum_{i=1}^N (\sigma^{(i)}_z+1) = \text{\# of spin-up sites}.
\end{equation}
$Q$ ranges from $0$ to $N$.
Suppose total spin number $Q$ is conserved, then
the system's Hilbert space factorizes into a direct sum of sectors with definite charge $Q$:
\begin{equation}
  \mathcal{H} = \bigoplus_{Q=0}^N H(Q).
\end{equation}
Given a pure state $\rho=\ket{\psi}\bra{\psi}$ for the total system in a specific $Q$ sector, we can divide the total system into two subsystems $A$ and $B$.
Because of the conserved charge, the total Hilbert space is no longer a direct tensor product of subspaces, $H(Q)\ne H_A \otimes H_B$.
Instead, for each $Q$ sector the Hilbert space has the following direct sum structure:
\begin{equation}
  H(Q) = \bigoplus_{q=0}^Q H(q) = \bigoplus_{q=0}^Q H_{A}(q) \otimes H_B(Q-q),
\end{equation}
where $q$ labels the amount of charge in subsystem $A$.
If $\rho$ has a definite charge, {\it i.e.}, $[\rho,\hat{Q}]=0$, it follows that the reduced density matrix $\rho_A$ also commutes with $\hat{Q}$ and is thus block-diagonal \footnote{To see why this is the case, we can write $\ket{\psi}=\sum_q \sqrt{p_q}\ket{\phi_q}$ with some $\ket{\phi_q}\in H(q)$. Then $\rho_A = \sum_{q,q'}\sqrt{p_qp_{q'}}\tr_B \ket{\phi_q}\bra{\phi_{q'}}=\sum_qp_q\tr_B\ket{\phi_q}\bra{\phi_q}$ since $\tr_B\ket{\phi_q}\bra{\phi_{q'}}\propto \delta_{qq'}$ from the definition of $\ket{q}$.}
\begin{equation}
  \label{blockdiag}
  \rho_A = \bigoplus_{q=0}^Q p_q \rho_{A}(q).
\end{equation}
$p_q$ can be thought of as the probability of finding $A$ with $q$ charges.
Given this decomposition, it is straightforward to write down the R\'enyi entropy
\begin{equation}
  S_n(\rho_A) = \frac{1}{1-n}\log \left[\sum_q p_q^n \tr (\rho^n_A(q))\right].
\end{equation}
The $n\to1$ limit gives the entanglement entropy 
\begin{equation}
  S(\rho_A) = \sum_q p_q S(\rho_A(q)) - \sum_q p_q \log p_q.
\end{equation}
We see that the entanglement entropy has two contributions: A weighted average of the entanglement entropy $S(\rho_A(q))=-\tr(\rho_A(q)\log \rho_A(q))$ for each subsector labeled by $q$\footnote{Precisely speaking, since only the total charge $Q$ is conserved, one should not regard $q$, the charge associated with subsystem A, as a label for sector. Here, we call $q$ as a label for \emph{subsector} in the sense that reduced density matrix for $\rho_A$ becomes block-diagonal as \eqref{blockdiag}.}, plus a Shannon term that accommodates for the uncertainty across different subsectors.

We now consider the state $\ket{\psi}$ to be drawn from a uniform Haar random ensemble within a $Q$ sector.
The Haar measure on $H(Q)$ can be factorized into products of Haar measures on each subsector:
\begin{align}
  \label{Qmeasure}
 d\mu_Q(\ket{\psi}) = d\nu(p_0,p_1,\cdots,p_Q) \prod_{q=0}^Q d\mu(\ket{\phi_q}),
\end{align}
where $\ket{\phi_q}$ represents uniform random states on $H(q)$ and $d\nu$ is given by the \emph{multivariate beta distribution}:
\begin{equation}
\label{betameasure}
  d\nu(p_0,p_1,\cdots,p_Q) = \frac{1}{\mathcal{Z}}\delta(\textstyle\sum_q p_q-1) \prod_q p_q^{d_q-1} dp_q,
\end{equation}
where $p_q$ are as defined in \eqref{blockdiag} and $d_q = |H(q)| = |H_{A}(q)| |H_B(Q-q)|$ is the dimension of the $q$-th subsector.
The normalization factor $\mathcal{Z}$ is fixed by requiring $\int d\nu=1$.
A derivation of \eqref{Qmeasure} can be found in Ref.~\cite{Bianchi:2019stn} and we reproduce it in Appendix~\ref{app:measure} for completeness.

The above discussion is rather abstract, so we demonstrate below with a simple concrete example. Consider a system of $N=4$ qubits. The system has five global charged sectors, labeled by a single number $Q$. Let's focus on the $Q=2$ sector here. If we pick $A$ to be the first two qubits and $B$ to be the rest two, then the Hilbert space further factorizes into local subsectors labeled by $q$, the number of charges in system $A$. The sector structure and its basis is illustrated as shown in Tab.~\ref{tab:sectors}.

\begin{table}[t]
\begin{tabular}{c c}
global sector & basis states \\
\hline
$Q=4$ & $\ket{\uparrow\uparrow\uparrow\uparrow}$ \\
$Q=3$ & $\ket{\uparrow\uparrow\uparrow\downarrow},\ket{\uparrow\uparrow\downarrow\uparrow}, \ket{\uparrow\downarrow\uparrow\uparrow}, \ket{\downarrow\uparrow\uparrow\uparrow}$ \\
$Q=2$ & $\ket{\uparrow\uparrow\downarrow\downarrow}, \ket{\uparrow\downarrow\uparrow\downarrow}, \ket{\uparrow\downarrow\downarrow\uparrow}, \ket{\downarrow\uparrow\downarrow\uparrow}, \ket{\downarrow\uparrow\uparrow\downarrow}, \ket{\downarrow\downarrow\uparrow\uparrow}$\\
$Q=1$ & $\ket{\uparrow\downarrow\downarrow\downarrow},\ket{\downarrow\uparrow\downarrow\downarrow}, \ket{\downarrow\downarrow\uparrow\downarrow}, \ket{\downarrow\downarrow\downarrow\uparrow}$ \\
$Q=0$ & $\ket{\downarrow\downarrow\downarrow\downarrow}$ \\
\\
local subsector & basis states \\
\hline
$q=2$ & $\ket{\uparrow\uparrow\downarrow\downarrow}$ \\
$q=1$ & $\ket{\uparrow\downarrow\uparrow\downarrow}, \ket{\uparrow\downarrow\downarrow\uparrow}, \ket{\downarrow\uparrow\downarrow\uparrow}, \ket{\downarrow\uparrow\uparrow\downarrow},$ \\
$q=0$ & $\ket{\downarrow\downarrow\uparrow\uparrow}$
\end{tabular}
\caption{(Top) The global $Q$ sectors for $N=4$ spins. (Bottom) The local subsectors within the global $Q=2$ sector.}
\label{tab:sectors}
\end{table}

A generic state in the $Q=2$ sector can be written as 
\begin{align}
\begin{split}
    \ket{\psi} &= \psi_{2} \ket{\uparrow\uparrow\downarrow\downarrow} + \psi_{1,1} \ket{\uparrow\downarrow\uparrow\downarrow} + \psi_{1,2} \ket{\downarrow\uparrow\uparrow\downarrow} \\
    & + \psi_{1,3} \ket{\uparrow\downarrow\downarrow\uparrow}+ \psi_{1,4} \ket{\downarrow\uparrow\downarrow\uparrow}+ \psi_{0} \ket{\downarrow\downarrow\uparrow\uparrow},
    \end{split}
\end{align}
where $\psi_{q,i}$ labels the basis coefficients in $H(q)$ (we ignore $i$ when $H(q)$ is one-dimensional).
The reduced density matrix on $A$ is given by $\rho_A=\tr_B(\ket{\psi}\bra{\psi})$:
\begin{align}
\begin{split}
\rho_A &= |\psi_2|^2 \ket{\uparrow\uparrow}\bra{\uparrow\uparrow} \\
&+ (\psi_{1,1}\ket{\uparrow\downarrow}+\psi_{1,2}\ket{\downarrow\uparrow})(\bar{\psi}_{1,1}\bra{\uparrow\downarrow}+\bar{\psi}_{1,2}\bra{\downarrow\uparrow})\\
&+ (\psi_{1,3}\ket{\uparrow\downarrow}+\psi_{1,4}\ket{\downarrow\uparrow})(\bar{\psi}_{1,3}\bra{\uparrow\downarrow}+\bar{\psi}_{1,4}\bra{\downarrow\uparrow})\\
&+ |\psi_0|^2\ket{\downarrow\downarrow}\bra{\downarrow\downarrow}.
\end{split}
\end{align}
We see that it indeed decomposes into subsector blocks labeled by $q$. The dimensions of the subsectors $d_q=|H(q)|$ are given by
\begin{equation}
    d_2=1, \quad d_1=4, \quad d_0=1.
\end{equation}
The uniform measure \eqref{Qmeasure} on the $Q=2$ sector decomposes into a product of the uniform measure over each of the $q$ subsectors $d\mu(\ket{\phi_q})$:
\begin{equation}
d\mu_{Q=2}(\ket{\psi})=d\nu(p_0,p_1,p_2)d\mu(\ket{\phi_0})d\mu(\ket{\phi_1})d\mu(\ket{\phi_2}).
\end{equation}
In the current example, the explicit forms for $d\mu(\ket{\phi_q})$ is given by
\begin{align}
\begin{split}
    d\mu(\ket{\phi_2})&=\frac{1}{2\pi}\delta(1-|\phi_2|^2)d\phi_{2} d\bar{\phi}_2  \\
    d\mu(\ket{\phi_1})&=\frac{1}{2\pi^2}\delta(1-\sum_{i=1}^{d_1}|\phi_{1,i}|^2)\prod_{i=1}^{d_1}d\phi_{1,i} d\bar{\phi}_{1,i}\\ 
    d\mu(\ket{\phi_0})&=\frac{1}{2\pi}\delta(1-|\phi_0|^2)d\phi_0 d\bar{\phi}_0.
\end{split}    
\end{align}
The coefficients $\phi_{q,i}$ are the rescaled version of $\psi_{q,i}$ such that $\sum_i|\phi_{q,i}|^2=1$.
The numbers $p_q$ satisfy $0\le p_q\le1$ and can be thought of as the probability to find the state in the $q$-th subsector.
The relative weights between each subsector is given by 
\begin{equation}
    d\nu(p_0,p_1,p_2) = \frac{1}{\mathcal{Z}}\delta(p_0+p_1+p_2-1) p_1^3dp_0dp_1dp_2.
\end{equation}

Back to the general discussion, the averaged R\'enyi entropy can be written as an integral over the uniform measure
\begin{align}
  S_n(\rho_A) 
  &= \frac{1}{1-n}\int d\mu_Q(\ket{\psi})\, 
     \ln \tr(\rho_A^n) \nonumber \\
  &= \frac{1}{1-n}\int d\nu(p_0,\cdots,p_Q)
     \prod_q d\mu(\ket{\phi_q}) \nonumber \\
  &\qquad \times   \ln\!\Big[\sum_q p_q^{\,n}
     \tr\!\big(\rho_A(q)^n\big)\Big].
\end{align}
We see that the problem of calculating the averaged R\'enyi entropy has been broken down into two parts: The R\'enyi entropy of a Haar random state within each $q$-subsector, and the weighted average over the subsector probabilities $p_q$.
In the von Neumann limit $n\to1$, we can simply apply the classic formula of Page \cite{Page:1993df}, 
\begin{equation}
  \label{Page}
  S(\rho_A(q)) = \Psi(d_A(q)d_B(q)+1)-\Psi(d_B(q)+1)-\frac{d_A(q)-1}{2d_B(q)},
\end{equation}
where $\Psi(x) = \Gamma'(x)/\Gamma(x)$ is the digamma function and we assume $d_A(q)\le d_B(q)$, where 
\begin{align}
d_A(q)\equiv \dim H_A(q) &\,, \quad 
d_B(q)\equiv \dim H_B(Q-q) \,, \\
 d_q &= d_A(q) d_B(q) \,.
\end{align}
For integer $n$ an explicit formula is also available but complicated \cite{Bianchi:2019stn}.
In the thermodynamic limit where $N,Q,q\gg 1$ while keeping the ratio $Q/N,q/N$ fixed, the dimensions of all the relevant Hilbert spaces are large. In this limit all the averaged R\'enyi entropies for a uniform random state coincide with the $n\to1$ limit \cite{Nadal:2010jbc}, where \eqref{Page} takes the simple form
\begin{equation}
  S_n(q) \approx  \min( \log d_A(q) , \log 
  d_B(q) ).
\end{equation}

For our setup of $N$ spins, the Hilbert space dimensions $d_A,d_B$ are given by
\begin{equation}
  d_A(q) = \binom{N_A}{q}, \quad d_B(q) = \binom{N_B}{Q-q},
\end{equation}
where $N_A,N_B$ are the number of sites in subsystem $A$ and $B$.
The measure for sector probabilities $d\nu(p_0,\cdots,p_q)$ also simplifies dramatically in the thermodynamic limit. One can show that $d\nu$ is sharply peaked around $q/N_A=(Q-q)/N_B$, which corresponds to the case where both subsystems have the same charge densities.
This suggests that to the leading order in $N$, the R\'enyi entropy receives dominant contribution from the subsector $q/N_A=Q/N$ only.
Thus we conclude that in the thermodynamic limit, the R\'enyi entropies have the following form
\begin{align}
  S_n(Q) &\approx \left(-n_Q\ln n_Q -(1-n_Q)\ln(1-n_Q)\right)\min(N_A,N_B)\nonumber \\
  &=S_Q\min(N_A,N_B),
\end{align}
where $n_Q \equiv Q/N$ is the averaged global charge density.
Therefore, we see that in the thermodynamic limit, apart from a prefactor $S_Q$, the shape of the Page curve remains unchanged.
$S_Q$ can be thought of as a proportionality factor that measures the relative size between a specific charged sector Hilbert space $H(Q)$ and the total Hilbert space $\mathcal{H}$.

\subsection{Genuine multi-entropy}
Here we briefly review the definition of (genuine) multi-entropies. For a more thorough exposition we refer the reader to Ref.~\cite{Iizuka:2025ioc} and \cite{Iizuka:2025caq}.
Suppose that we divide the total system $M$ into $\mathtt{k}$ subsystems $R_1,R_2,\cdots,R_{\mathtt{k}}$, under which the total Hilbert space factorizes as $\mathcal{H} = H_1 \otimes H_2 \otimes \cdots \otimes H_{\mathtt{k}}$.
A pure state $\ket{\psi}$ on $M$  can be thought of as a rank-$\mathtt{k}$ tensor:
\begin{equation}
  \ket{\psi} = \sum_{i_1=1}^{d_1}\sum_{i_2=1}^{d_2}\cdots\sum_{i_\mathtt{k}=1}^{d_\mathtt{k}}   \psi_{i_1 i_2\cdots i_{\mathtt{k}}}\ket{i_1}\ket{i_2}\cdots \ket{i_{\mathtt{k}}}.
\end{equation}
Given an integer $n>1$, the $n$-th \emph{R\'enyi multi-entropy} on a $\mathtt{k}$-partite system is defined as the symmetric contraction \cite{Gadde:2022cqi}
\begin{align}
  \label{def_ME}
  \S[k]_n &= \frac{1}{1-n}\frac{1}{n^{\mathtt{k}-2}}\log \left(\frac{\Z[k]_n}{(\Z[k]_1)^{n^{\mathtt{k}-1}}}\right), \\
  \Z[k]_n &= \bra{\psi}^{\otimes n^{\mathtt{k}-1}} \Sigma_1(g_1) \Sigma_2(g_2)\cdots \Sigma_\mathtt{k}(g_\mathtt{k}) \ket{\psi}^{\otimes n^{\mathtt{k}-1}},
\end{align}
where $\Sigma_\mathtt{i}$ are ``twist operators'' that permute the $\mathtt{i}$-th indices of $\ket{\psi}^{\otimes n^{\mathtt{k}-1}}$ according the permutation group element $g_\mathtt{i}\in S_{n^{\mathtt{k}-1}}$. The detailed forms of $g_\mathtt{i}$ are specified by their action on the $n^{\mathtt{k}-1}$ dimensional hypercube lattice of length $n$:
\begin{equation}
  g_\mathtt{i} \cdot (x_1,\cdots, x_\mathtt{i},\cdots,x_{\mathtt{k}-1}) = (x_1,\cdots, x_\mathtt{i}+1,\cdots,x_{\mathtt{k}-1})
\end{equation}
for $1\le\mathtt{i}\le \mathtt{k}-1$, and $g_{\mathtt{k}}=e$. The list $(x_1,\cdots,x_{\mathtt{k}-1})$ labels integer lattice points on the hypercube with periodic boundary conditions where we identify $x_{\mathtt{i}}=n+1$ with $x_{\mathtt{i}}=1$.

\begin{figure}[t!]
  \centering
  \includegraphics[width=\linewidth]{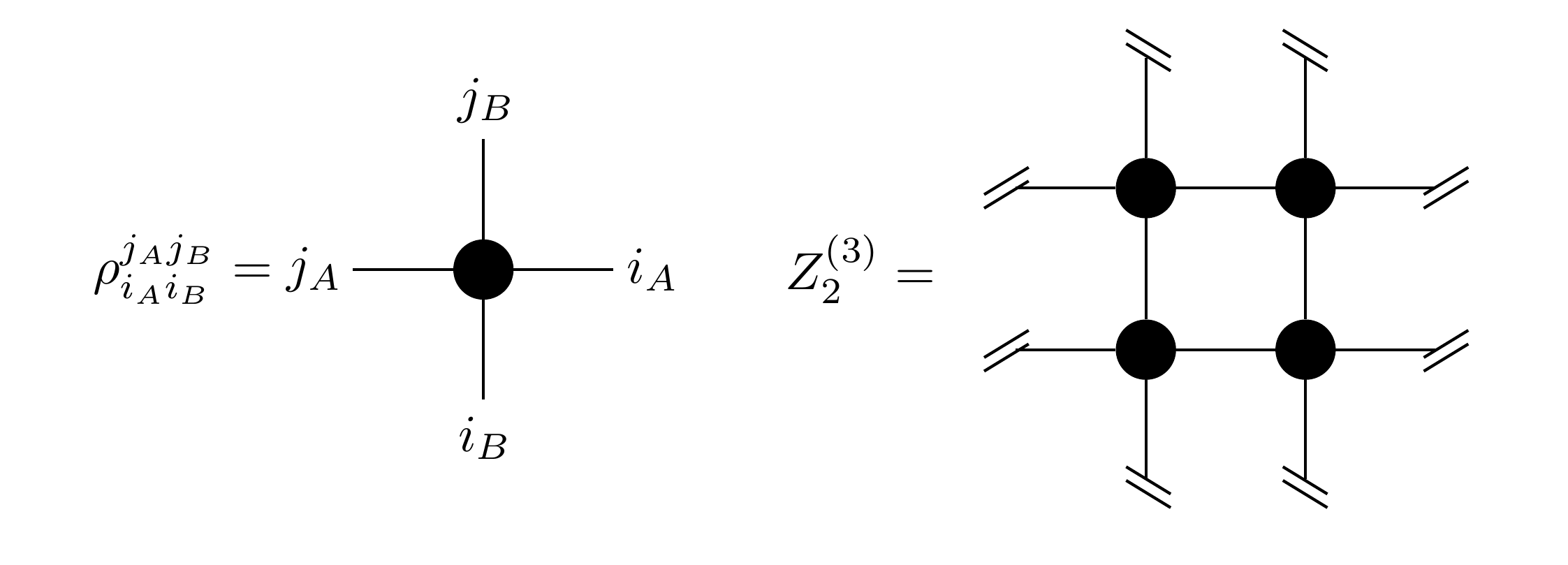}
  \caption{The contraction pattern for the reduced density matrix $\rho_{AB}=\tr_C(\ket{\psi}\bra{\psi})$ used to construct the tripartite R\'enyi multi-entropy $\S[3]_2$. Figure taken from Ref.~\cite{Iizuka:2025ioc}.}
  \label{Z23}
\end{figure}

The replica partition function $\Z[k]_n$ can be written as a symmetric contraction over $n^{\mathtt{k}-1}$ reduced density matrices. We illustrate this using two examples. We set $n=2$ for simplicity. If $\mathtt{k}=2$, see that $\Z[2]_2$ reduces to the square of the reduced density matrix $\rho_A=\tr_{B}\ket{\psi}\bra{\psi}$:
\begin{equation}
    \Z[2]_2=(\rho_A)^{i_2}_{i_1}(\rho_A)^{i_1}_{i_2}=\tr(\rho_A^2).
\end{equation}
For integer $n$ one can show that $\Z[2]_n=\tr(\rho_A^n)$. Therefore the $\mathtt{k}=2$ R\'enyi multi-entropy reduces to the usual R\'enyi entropy $S_n(A)$.
For $\mathtt{k}=3$, let $\rho_{AB}=\tr_C(\ket{\psi}\bra{\psi})$ be the reduced density matrix obtained by tracing out the third subsystem. $\Z[3]_2$ can be written as a contraction over $2^2=4$ copies of $\rho_{AB}$:
\begin{equation}
Z^{(\mathtt{3})}_n = (\rho_{AB})^{i_2j_3}_{i_1j_2}(\rho_{AB})^{i_1j_4}_{i_2j_2}(\rho_{AB})^{i_4j_1}_{i_3j_3}(\rho_{AB})^{i_3j_2}_{i_4j_4},
\end{equation}
which can be thought of as a contraction over a square lattice of reduced density matrices of length $n=2$, as shown in Fig.~\ref{Z23}.
For higher $\mathtt{k}$, $\Z[k]_n$ can be constructed similarly by consider a $\mathtt{k}-1$ dimensional hypercube lattice of $\mathtt{k}-1$-partite reduced density matrices $\rho_{R_1R_2\cdots R_{\mathtt{k}-1}}$ with length $n$, supplied with periodic boundary conditions in each direction.

The $\mathtt{k}$-partite multi-entropy $\S[k]$ is defined in the usual way by analytically continuing $n\to1$ in \eqref{def_ME}. However in this letter we will only work with multi-entropies with integer R\'enyi indices, without worrying about the continuation $n\to 1$.

Multi-entropies enjoy properties that a multi-partite entanglement measure is expected to satisfy, such as invariance under the action of local unitaries and monotonicity under coarse-graining \cite{Gadde:2024jfi,Gadde:2023zni}. However, it is not very selective in the sense that the $\mathtt{k}$-partite multi-entropy $\S[k]_n$ is also sensitive all the lower $\tilde{\mathtt{k}}$-partite entanglements with $\tilde{\mathtt{k}}\le \mathtt{k}$.
To overcome this issue, one can construct new multi-partite entanglement signals\footnote{We use the word ``signal'' here since while a non-zero $\GM[k]_n$ implies the existence of multi-partite entanglement, $\GM[k]_n$ by itself is in general not positive definite.} by starting with $\S[k]$ and systematically subtracting all the lower-partite contributions. The resulting quantities are linear combinations of multi-entropies, known as \emph{Genuine multi-entropies} $\GM[k]_n$ \cite{Iizuka:2025ioc,Iizuka:2025caq}.

The genuine multi-entropies $\GM[k]_n$ satisfy the following key properties:
\begin{itemize}
\item $\GM[k]_n$ contains the $\mathtt{k}$-partite (R\'enyi) multi-entropy $\S[k]_n$.
\item $\GM[k]_n$ vanishes for all $\tilde{\mathtt{k}}$-partite entangled states with $\tilde{\mathtt{k}}\le \mathtt{k}$. \footnote{These states include but are not limited to factorized states. For example, the triangle state $\ket{\psi}_{ABC}=\ket{\psi_{A_1,B_2}}\ket{\psi_{B_1,C_2}}\ket{\psi_{C_1,A_2}}$ with $A=A_1\cup A_2, B=B_1\cup B_2, C=C_1\cup C_2$ is not factorizable but built from tensor products of bipartite states only, and is not tripartitely entangled..}
\end{itemize}
In general, there will be more than one linear combinations of $\S[k]$ which satisfy the above criteria. This is related to the fact that, as opposed to the bipartite case, there are more than one inequivalent multi-partite entangled states which cannot be converted to one another by LOCC  (Local Operations and Classical Communication).
The explicit formulae of GM for small $\mathtt{k}$ have been worked out in Refs.~\cite{Iizuka:2025ioc,Iizuka:2025caq}. In this letter we will focus on the tripartite and quadripartite case:
\onecolumngrid
  \begin{align}
    \label{GM3}
    & \,\,\,\, \quad \GM[3]_n(A:B:C) = \S[3]_n(A:B:C) - \frac{1}{2}\left(\S[2]_n(A)+\S[2]_n(B)+\S[2]_n(C)\right) \,, \\
     \label{GM4}
\begin{split}
& \GM[4]_n(A:B:C:D) =  S_n^{(4)}(A:B:C:D) \\
& \qquad -\frac{1}{3}\Big(S_n^{(3)}(AB:C:D)+S_n^{(3)}(AC:B:D)
+S_n^{(3)}(AD:B:C)\\
& \qquad\;\;\;\;\;\;\;+S_n^{(3)}(BC:A:D)+S_n^{(3)}(BD:A:C)+S_n^{(3)}(CD:A:B)\Big) \\
& \qquad+a\left(S_n^{(2)}(AB:CD)+S_n^{(2)}(AC:BD)+S_n^{(2)}(AD:BC)\right)  \\
& \qquad+\left(\frac{1}{3}-a\right)\left(S_n^{(2)}(ABC:D)+S_n^{(2)}(ABD:C)+S_n^{(2)}(ACD:B)+S_n^{(2)}(BCD:A)\right).
\end{split}
\end{align}
\twocolumngrid
The $\mathtt{k}=3$ quantity \eqref{GM3} has been considered in \cite{Penington:2022dhr,Gadde:2023zzj,Harper:2024ker,Liu:2024ulq} as well.
The $\mathtt{k}=4$ quantity \eqref{GM4} includes one real free parameter $a$. We can recast it in a more suggestive form using
\begin{align}
  \begin{split}
  &\GM[4]_n(A:B:C:D) = \\
  &\quad \GM[4]_n(A:B:C:D)|_{a=1/3} - aI_{3,n}(A:B:C)
  \end{split}
\end{align}
where
\begin{align}
  \begin{split}
    &I_{3,n}(A:B:C) = S_n(A)+S_n(B)+S_n(C)\\
  &\quad -S_n(AB)-S_n(BC)-S_n(AC)+S_n(ABC)
  \end{split}  
\end{align}
is the (R\'enyi) tripartite information, which has been shown to be a signal for genuine quadripartite entanglement independently \cite{Balasubramanian:2014hda}.
Since $I_3$ are obtained by linear combinations of bipartite entanglement entropies, in this letter we will be focusing on the $a=1/3$ quantity defined in \eqref{GM4}. In the remaining sections we will implicitly assume $a=1/3$ whenever we refer to $\GM[4]_n$. We should stress that there is nothing special about this particular value.
We choose $a = 1/3$ simply to illustrate one diagnostic feature of the quadripartite genuine multi-entropy.

Before concluding, we emphasize that the construction of $\GM[k]$ can be straightforwardly extended to all integer $\mathtt{k}$. For explicite formula of $\GM[\mathtt{k}]$ with $\mathtt{k} = 5$, see \cite{Iizuka:2025caq}. This is a significant advantage for genuine multi-entropy as a multi-partite entanglement signal compared to other proposed measures.

\section{Symmetry-resolved multi-entropy curves}
\label{sec:SRGM}
In this section, we study the symmetry resolution of GM,  especially symmetry resolution of the black hole (genuine) multi-entropy curves. In Sec.~\ref{sec:Haar} we study Haar random states and provide analytical formulas for both the multi-entropy and GM in the thermodynamic limit. We compare our results with numerical evaluations on random uniform state on a chain of $N=12$ spins. In Sec.~\ref{sec:graph} we study random graph states. We give explicit analytical formulas for the (non-symmetry-resolved) GM and numerical results for symmetry-resolved GM. Although our results are mostly numerical, we find that the multi-partite entanglement structure for graph states seem to obey the same universality after projection to a specific charged sector.

We briefly recall the definition of black hole (genuine) multi-entropy curves. The idea is to model the total system (an evaporating black hole and its Hawking radiation) by a state drawn from some random ensemble.  We divide the entire Hawking radiation into $\mathtt{k}-1$ subsystems of identical Hilbert space dimension,
\begin{equation}
  d_{\mathrm{R1}} = \cdots = d_{\mathrm{R}(\mathtt{k}-1)} \equiv d_{\mathrm{R}},
\end{equation}
where $d_{{\rm R}i}$ denotes the dimension of the $i$-th radiation subsystem $(i = 1, \cdots, \mathtt{k}-1)$.  
One can, for example, think of a spherically symmetric black hole emitting Hawking radiation that is divided 
into $\mathtt{k}-1$ angular sectors, each corresponding to a radiation subsystem.  
In this picture, the assumption of identical Hilbert space dimensions simply reflects the symmetry among these angular sectors. 
We also fix the total Hilbert space dimension as
\begin{equation}
    d_{\rm Total} = d_{\rm R}^{\,\mathtt{k}-1} d_{\rm BH} = \text{const.},
\end{equation}
where $d_{\rm BH}$ represents the dimension of the evaporating black hole. The black hole (genuine) multi-entropy curves are then obtained by plotting the (genuine) R\'enyi multi-entropy 
while varying $d_{\rm R}$ from $d_{\rm R}=1$ (the beginning of the evaporation process) to $d_{\rm BH}=1$ (the endpoint of evaporation).  
See Fig.~2 in~\cite{Iizuka:2024pzm} for the multi-entropy curves and Fig.~3 in~\cite{Iizuka:2025caq} 
for the genuine multi-entropy curves.  
In particular, for $\mathtt{k}=2$, this construction reduces to the usual Page curve, 
which describes the bipartite entanglement entropy between the black hole and its Hawking radiation.

\subsection{Haar random states}
\label{sec:Haar}
\subsubsection{Tripartite multi-entropy}
We first work out the symmetry-resolved multi-entropy for the tripartite case before generalizing to arbitrary $\mathtt{k}$.
For $\mathtt{k}=3$ there is a similar decomposition of the Hilbert space into subsystem charges, and the only difference to the bipartite case is that the sectors are labeled by two numbers $(q_A,q_B)$ instead of one:
\begin{align}
     H(Q) &= \bigoplus_{q_A,q_B}H(q_A,q_B,q_C) \\
  &=\bigoplus_{q_A,q_B} H_A(q_A) \otimes H_B(q_B) \otimes H_C(q_C), \nonumber
\end{align}
where $q_C=Q-q_A-q_B$.
The uniform measure on $H(Q)$ factorizes in a similar manner \footnote{To see why, note that the factorization of the measure $d\mu_Q(\ket{\psi})$ only depends on the dimensions of symmetry sectors in the Hilbert space. Thus we can simply reuse \eqref{Qmeasure}-\eqref{betameasure} and replace $d_q$ accordingly.}
\begin{equation}
\label{Qmeasure2}
 d\mu_Q(\ket{\psi}) = d\nu(\{p_{q_Aq_B}\}) \prod_{q_A,q_B} d\mu(\ket{\phi_{q_Aq_B}}),
\end{equation}
where $d\mu(\ket{\phi_{q_Aq_B}})$ is the uniform measure on $H(q_A,q_B)$ and
\begin{equation}
  d\nu(\{p_{q_Aq_B}\}) = \frac{1}{\mathcal{Z}}\delta\bigg(\sum_{q_A,q_B}p_{q_Aq_B}-1\bigg) \prod_{q_A,q_B} p_{q_Aq_B}^{d_{q_Aq_B}-1} dp_{q_Aq_B},
\end{equation}
where $d_{q_Aq_B}$ is the dimension of the Hilbert space in the sector labeled by $(q_A,q_B)$ and it obeys $d_{q_Aq_B}=d_Ad_Bd_C$, where $d_A \equiv \dim H_A(q_A)$, 
$d_B \equiv \dim H_B(q_B)$,  $d_C \equiv \dim H_C(q_C)$.

Similar to the bipartite case we also demonstrate \eqref{Qmeasure2} with a simple example. Let's consider the system of $N=3$ qubits and we pick the subsystem $A,B,C$ to be the first, second and third qubit respectively. 
The sectors are symmetric under a $\mathbb{Z}_2$ action which flips all spins, thus it suffices to look at $Q=0$ and $Q=1$. Since the $Q=0$ sector is trivial, let's focus on $Q=1$. There are three states in this sector: \(\ket{\uparrow\downarrow\downarrow}\), \(\ket{\downarrow\uparrow\downarrow}\) and \(\ket{\downarrow\downarrow\uparrow}\). Each of them belongs to a subsector labeled by two numbers $(q_A,q_B)$, namely the amount of charges in system $A$ and $B$, as illustrated in Tab.~\ref{tab:sectors2}.
\begin{table}[t]
\begin{tabular}{c c}
\begin{tabular}[t]{c c}
global sector & basis \\
\hline
$Q=3$ & $\ket{\uparrow\uparrow\uparrow}$ \\
$Q=2$ & $\ket{\uparrow\uparrow\downarrow},\ket{\uparrow\downarrow\uparrow}, \ket{\downarrow\uparrow\uparrow}$ \\
$Q=1$ & $\ket{\uparrow\downarrow\downarrow}, \ket{\downarrow\uparrow\downarrow}, \ket{\downarrow\downarrow\uparrow}$\\
$Q=0$ & $\ket{\downarrow\downarrow\downarrow}$ \\
\end{tabular} \quad &
\begin{tabular}[t]{c c}
local subsector & basis \\
\hline
$(q_A,q_B)=(1,0)$ & $\ket{\uparrow\downarrow\downarrow}$ \\
$(q_A,q_B)=(0,1)$ & $\ket{\downarrow\uparrow\downarrow}$ \\
$(q_A,q_B)=(0,0)$ & $\ket{\downarrow\downarrow\uparrow}$
\end{tabular}
\end{tabular}
\caption{(Left) The global $Q$ sectors for $N=3$ spins. (Right) The local tripartite subsectors within the global $Q=1$ sector.}
\label{tab:sectors2}
\end{table}

Since each of the subsector is one dimensional, $d_{10}=d_{01}=d_{00}=1$,
the integral measure is simply 
\begin{equation}
d\mu_{Q=1}(\ket{\psi})  = \frac{1}{\mathcal{Z}}\delta(p_{10}+p_{01}+p_{00}-1) dp_{10}dp_{01}dp_{00} 
\end{equation}

For a spin chain of $N$ sites and total charge $Q$ the dimensions $q_A,q_B,q_C$ are given by
\begin{align}
  d_A = \binom{N_A}{q_A},\quad d_B = \binom{N_B}{q_B},\quad d_C = \binom{N_C}{Q-q_A-q_B}.
\end{align}
where $N_C := N-N_A-N_B$. Again, same as in the bipartite case, in the thermodynamic limit where $d_A,d_B,d_C\gg1$, one can show that the measure $d\nu$ is sharply peaked around $q_A/N_A=q_B/N_B=Q/N$, {\it i.e.}, when the charges are uniformly distributed in the total system.
Consequently, to the leading order in $N$, the symmetry-resolved multi-entropy $\S[3]_n$ is identical to the multi-entropy on a tripartite random state with dimensions $(d_A,d_B,d_C)$.

\begin{figure}[t]
\centering
\includegraphics[width=\linewidth]{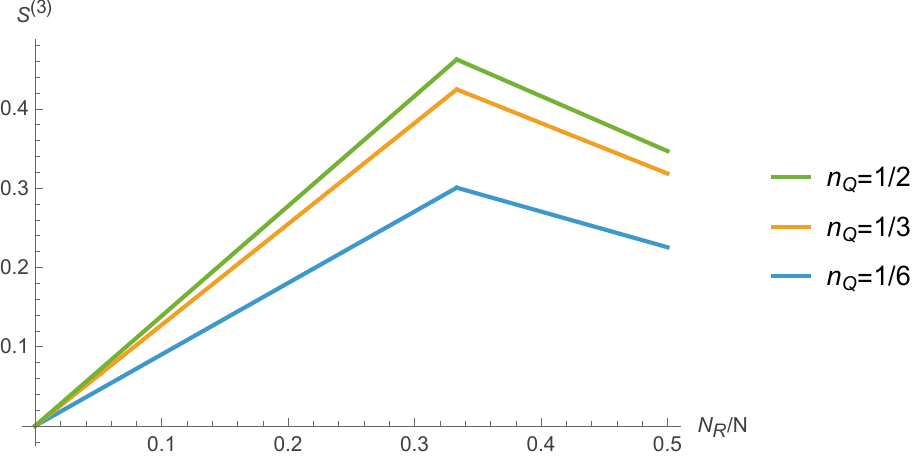}
\caption{A plot of the tripartite multi-entropy curves \eqref{MES3} for different global charge densities $n_Q$.}
\label{fig:MES3}
\end{figure}

The multi-entropy for random states of large dimensions has already been studied in Ref.~\cite{Iizuka:2024pzm} and we can simply reuse their results. Here we will set $N_A=N_B=N_R$ and $N_C=N_{BH}$. 
The tripartite multi-entropy for large-dimensional Haar random states was conjectured to be
\begin{equation}
\label{S3conj}
  \S[3]_n(d_R,d_R,d_{BH}) \approx \log(d_{total})-\max(\log(d_R),\log(d_{BH})).
\end{equation}
Note that it is independent of the R\'enyi index $n$.
Using this expression and the asymptotic form of binomial coefficients we find that
\begin{equation}
\label{MES3}
\S[3]_n(N;Q;N_R) =
\begin{cases}
2S_QN_R, \quad &N_R < N/3, \\
S_Q(N-N_R), \quad &N_R > N/3,
\end{cases}
\end{equation}
where the prefactor $S_Q$ takes the same form as in the bipartite case:
\begin{equation}
  S_Q \equiv -n_Q\log n_Q- (1-n_Q)\log(1-n_Q),\quad n_Q = \frac{Q}{N}.
\end{equation}
We emphasize that the appearance of the same proportionality factor $S_Q$ here is a direct consequence of \eqref{S3conj}, which is not apparent a priori. We plot \eqref{MES3} in Fig.~\ref{fig:MES3} for illustration.

To summarize, we find that to the leading order in $N$ in the thermodynamic limit, the multi-entropy of a charged sector $Q$ is simply the multi-entropy of the fully random state (without any conserved charge), multiplied by a scaling factor $S_Q$ that accounts for the relative dimension between the sector and the full Hilbert space. 

\subsubsection{$\mathtt{k}$-partite multi-entropy}
The higher-partite cases are straightforward generalizations of $\mathtt{k}=3$. The Hilbert space of global charge $Q$ has the following decomposition
\begin{equation}
  H(Q) = \bigoplus_{q_1,\cdots,q_k} H_1(q_1) \otimes \cdots \otimes H_k(q_k),\quad Q = \sum_k q_k,
\end{equation}
where $q_\mathtt{i}$ labels the amount of charges in the $\mathtt{i}$-th subsystem.
In the thermodynamic limit, the system is again dominated by the uniform charge distribution (as long as $\mathtt{k}\ll N$)
\begin{equation}
  \frac{q_1}{N_1} = \frac{q_2}{N_2} = \cdots = \frac{q_\mathtt{k}}{N_\mathtt{k}} = \frac{Q}{N}.
\end{equation}
And thus we have
\begin{equation}
  \S[k](Q;N_\mathtt{i}) \approx \S[3]\left(d_1,d_2, \cdots ,d_\mathtt{k}\right),\quad d_\mathtt{i} = \binom{N_\mathtt{i}}{q_\mathtt{i}}.
\end{equation}
Similarly, by assuming $N_1=N_2=\cdots=N_{\mathtt{k}-1}=N_R$ and using the conjectured results in Ref.~\cite{Iizuka:2024pzm} as well as the asymptotic form of binomial coefficients, we arrive at the simple formula
\begin{equation}
  \label{MESk}
\S[k](N;Q;N_R) =
\begin{cases}
S_Q(k-1)N_R, \quad &N_R < N/k \\
S_Q(N-N_R), \quad &N_R > N/k
\end{cases}
\end{equation}
as the final expression. \eqref{MESk} is one of the main results of this letter.

\subsubsection{Genuine multi-entropies}
\begin{figure}[t]
\centering
\includegraphics[width=\linewidth]{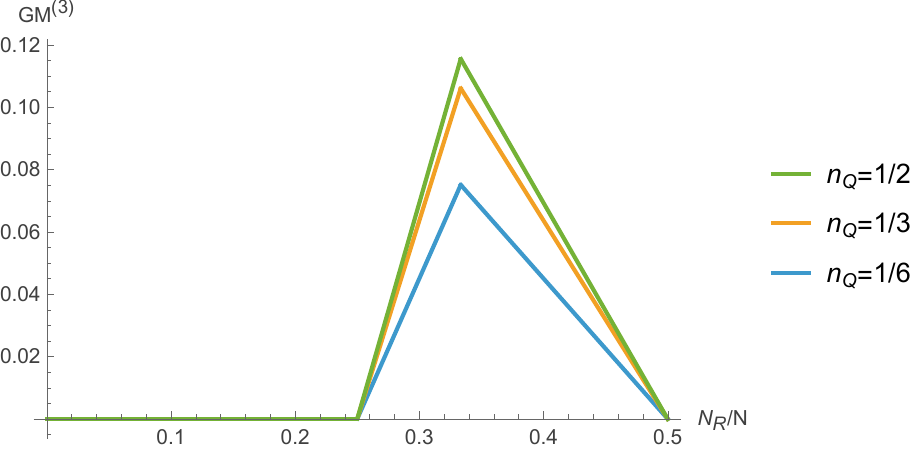}
\caption{A plot of the tripartite genuine multi-entropy \eqref{GME3} curves for different global charge densities $n_Q$. The quadripartite genuine multi-entropy \eqref{GME4} also follows a similar curve.}
\label{fig:GME3}
\end{figure}

Since we have obtained the expression \eqref{MESk} for symmetry-resolved multi-entropies under the thermodynamic limit, it is a straightforward exercise to plug in these expressions into the formulae of GM \eqref{GM3},\eqref{GM4} to find out the expression for symmetry-resolved genuine GM. We report the expressions for $\mathtt{k}=3,4$ here:
  \begin{align}
  \label{GME3}
    \GM[3](N;Q;N_R)=
    \begin{cases}
  0,\quad &N_R\le N/4, \\
  S_Q(2N_R-N/2),\quad & N/4 <N_R\le N/3, \\
  S_Q(N/2-N_R), \quad & N_R> N/3.
  \end{cases}
  \end{align}
  
  \begin{align}
  \label{GME4}
    \GM[4](N;Q;N_R) =
    \begin{cases}
  0,\quad &N_R\le N/5, \\
  S_Q(5N_R-N),\quad &N/5<N_R\le N/4, \\
  S_Q(N-3N_R), \quad &N_R>N/4.
  \end{cases}
  \end{align}
  
As a record here we plot the $\mathtt{k}=3$ genuine multi-entropy \eqref{GME3} in Fig.~\ref{fig:GME3} for illustration.

\subsubsection{Numerics}
\begin{figure*}[t]
  \centering
  \includegraphics[width=.4\linewidth]{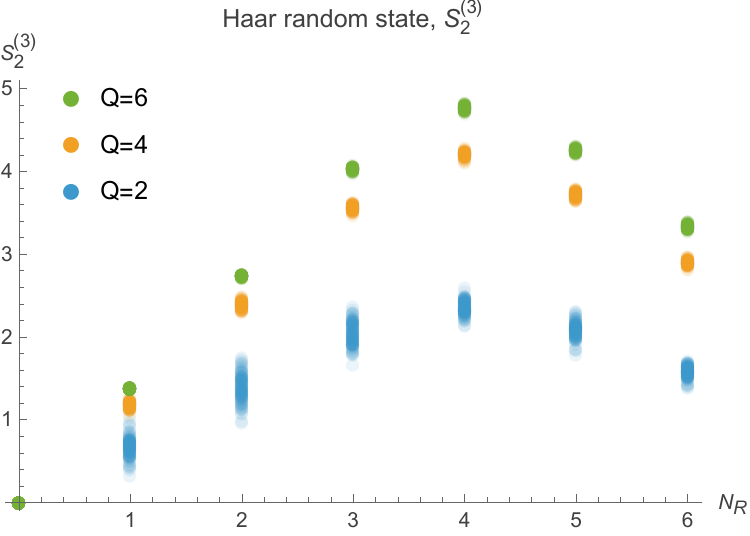}
  \hspace{5mm}
  \includegraphics[width=.4\linewidth]{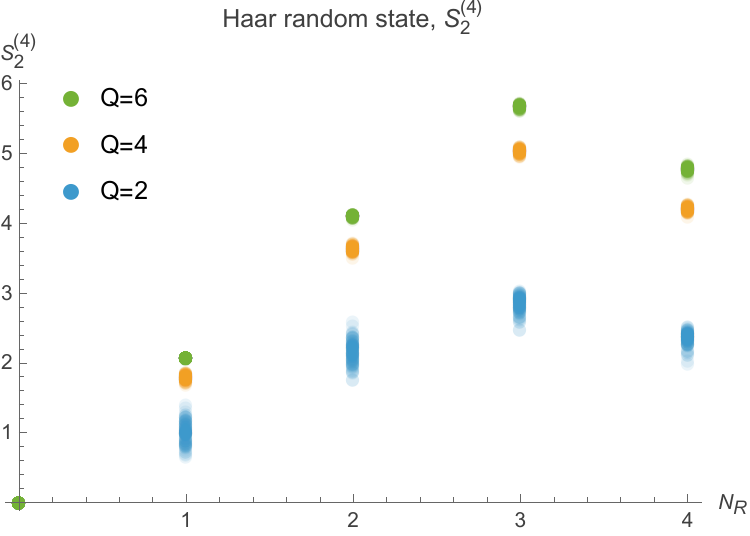}
  \includegraphics[width=.4\linewidth]{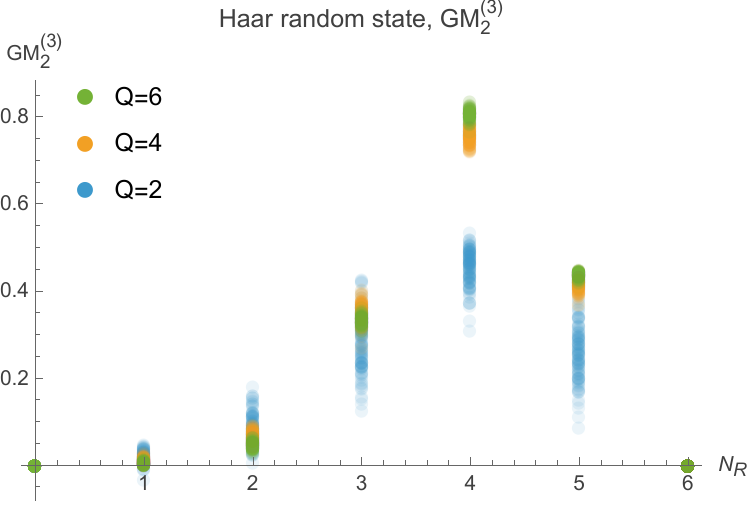}
    \hspace{5mm}
  \includegraphics[width=.4\linewidth]{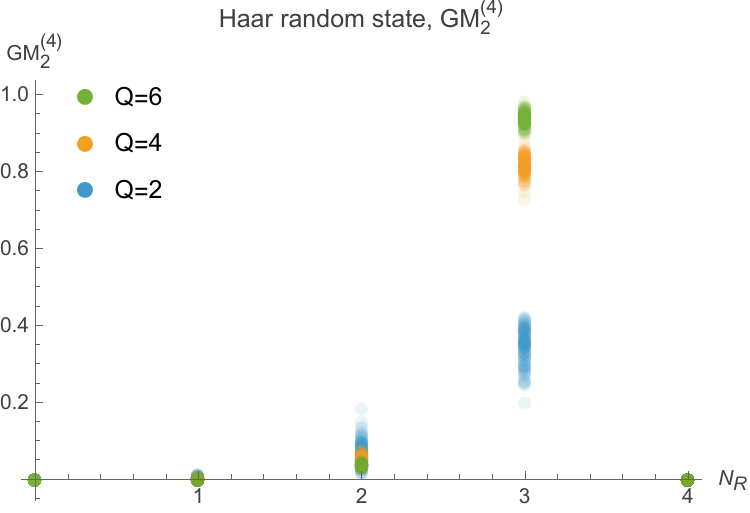}
  \caption{Numerical evaluation for the symmetry-resolved $\mathtt{k}=3$ and $\mathtt{k}=4$ (genuine) $n=2$ multi-entropy curves on $N=12$ qubits with global charge $Q=2,4,6$.  The top row shows multi-entropies and the bottom row shows genuine multi-entropies. For each $(N_R,Q)$ point we sample $100$ random vectors from the uniform Haar ensemble within the sector Hilbert space. The darkness of the data points represents the number of samples (darker = more). The large variance and the smoothed out transitions ({\it i.e.},  $N_R=N/\mathtt{k}$ for both $\S[k]$ and $\GM[k]$, as well as $N_R=2$ for $\GM[k]$) are expected from finite dimensional corrections to the analytical results in the thermodynamic limit.}
  \label{fig:Haar}
\end{figure*}

So far our focus has been on the behavior of GM in the thermodynamic limit.
Here we perform a numerical study of finite $N$ cases. In particular, we numerically evaluated the symmetry-resolved (genuine) multi-entropy for uniform random states on a chain of 12 spins.
We evaluate the $\mathtt{k}=3$ and $\mathtt{k}=4$ symmetry-resolved multi-entropies and GM for R\'enyi index $n=2$, projected down to global charged sectors of $Q=2,4,6$ respectively. The results are shown in Fig.~\ref{fig:Haar}. The results show similar trends to the analytic formulae we derived in the previous subsection. 

There is one caveat to note. For finite $N$, the genuine multi-entropy becomes slightly nonzero even before the Page time, unlike in the thermodynamic limit where there is a sharp transition. However, if one considers the dependence over $N_R$ for some fixed $Q$, one can see that as $Q$ increases, the curve asymptotically approaches the genuine multi-entropy curves shown in Fig.~\ref{fig:GME3}. This is because the accessible states in each sector increases as $Q$ becomes larges (as long as $Q<N/2$).

\onecolumngrid

\begin{center}
  \rule{\textwidth}{0.3pt}
\end{center}
\twocolumngrid


\subsection{Graph states}
\label{sec:graph}

We now shift our focus and investigate the multi-partite entanglement structure of an important class of states essential to quantum computation known as \emph{graph states}\footnote{For a detailed review on graph states and their entanglement properties we refer the reader to Ref.~\cite{Hein:2006uvf}}.
A graph state is specified by a graph $G=\{V,E\}$. Each vertex of the graph corresponds to a qubit, which is initialized in the Hadamard states $\ket{+} = \frac{1}{\sqrt{2}}(\ket{\uparrow}+\ket{\downarrow})$.
For each edge connecting two vertices $(v_1,v_2)$ we apply the controlled $Z$ gate $U^{(v_1,v_2)}$ on the two qubits:
\begin{equation}
  \ket{G} = \prod_{(v_1,v_2)\in E} U^{(v_1,v_2)}\ket{+}^{\otimes V},
\end{equation}
where
\begin{equation}
  U^{(v_1,v_2)} =
  \begin{pmatrix} 1 & 0 & 0 & 0 \\ 0 & 1 & 0 & 0 \\ 0 & 0 & 1 & 0 \\ 0 & 0 & 0 & -1 \end{pmatrix}
\end{equation}
in the two qubit basis of $(v_1,v_2)$. 
Since all controlled-$Z$ gates are diagonal in the computational basis, clearly they commute, so the order in $\prod_{(v_1,v_2)\in E}U^{(v_1,v_2)}$ does not matter. 

To build intuition, consider a single edge $(v_1,v_2)$.  Note that for a single edge $(v_1,v_2)$, starting from $\ket{+}\otimes\ket{+}$, a controlled-$Z$ gate $U$ produces
\begin{align}
U\ket{++}
&= \tfrac{1}{2}\!\left(\ket{\uparrow \uparrow}+\ket{\uparrow \downarrow}+\ket{\downarrow \uparrow}-\ket{\downarrow \downarrow}\right) \\
&= (1 \,\!\otimes H)\, \tfrac{1}{\sqrt{2}}\!\left(\ket{\uparrow \uparrow}+\ket{\downarrow \downarrow}\right) \,,
\end{align}
where $H$ is the Hadamard gate, which maps the $Z$-eigenbasis to the $X$-eigenbasis as 
\begin{align}
H\ket{\uparrow}=\frac{\ket{\uparrow}+\ket{\downarrow}}{\sqrt2}\,, \quad 
H\ket{\downarrow}=\frac{\ket{\uparrow}-\ket{\downarrow}}{\sqrt2}.
\end{align}
Thus, $U\ket{++}$ is locally unitary equivalent to a Bell pair and carries one bit of entanglement. 

All graph states are stabilizer states since they are built from Hadamard and controlled Pauli gates. In fact there is an equivalent way of defining graph states using the stabilizer formalism: For each $v\in V$ we define
\begin{equation}
  S_v = \sigma_x^{(v)}\prod_{u\in N(v)}\sigma_z^{(u)},
\end{equation}
where $N(v)$ denotes the neighboring vertices of $v$ in $G$. Then all $S_v$ commute and $\ket{G}$ is the unique simultaneous eigenstate with eigenvalue $+1$ for all $S_v$.

Stabilizer states have a very specific entanglement structure. Let $\mathcal{S}$ be the stabilizer group of the state $\ket{\psi}$, and $\mathcal{S}_A(\mathcal{S}_B)$ be the subgroup of $\mathcal{S}$ that acts as identity on $B$ and $A$, respectively. The R\'enyi entropies of $\ket{\psi}$ are independent of $n$ and are given by \cite{Hein:2004zjp,Fattal:2004frh,Hein:2006uvf}
\begin{equation}
\label{stabilizerEE}
  S_n(A) = \frac{1}{2}\log\left(\frac{|\mathcal{S}|}{|\Braket{\mathcal{S}_A,\mathcal{S}_B}|}\right) = \frac{1}{2}\log\left(\frac{|\mathcal{S}|}{|\mathcal{S}_A||\mathcal{S}_B|}\right),
\end{equation}
where $|\mathcal{S}|$ denotes the cardinality of $\mathcal{S}$ and $\Braket{\mathcal{G},\mathcal{H}}$ denotes the group generated by $\mathcal{G}$ and $\mathcal{H}$. 
For a graph state $\ket{G}$, let $\Gamma$ be the adjacency matrix of $G$. $\Gamma$ has the form
\begin{equation}
  \Gamma =
  \begin{pmatrix}
    \Gamma_A & \Gamma^T_{AB} \\ \Gamma_{AB} & \Gamma_B
  \end{pmatrix},  
\end{equation}
where $\Gamma_{A} (\Gamma_B)$ represents the edges within $A (B)$ and $\Gamma_{AB}$ represents the edges that connect between $A$ and $B$.
The R\'enyi entropies of $\ket{G}$ are given by the $\mathbb{Z}_2$ rank of $\Gamma_{AB}$ \cite{Hein:2004zjp,Fattal:2004frh}:
\begin{equation}
  \label{graphEE}
  S_n(A) = \text{rank}_{\mathbb{Z}_2}(\Gamma_{AB})\log 2.
\end{equation}
Thus $S_n(A)$ can only take values in integer multiples of $\log 2$.

In the following we numerically compute both the unresolved and symmetry-resolved (genuine) multi-entropies for random graph states. For better comparison with our numerics on Haar random states, we choose to work with random graph states with $N=12$ qubits. 
The graphs used to construct the states are sampled from the Erd\H{o}s--R\'enyi (Bernoulli) random graph model $G(N,p)$  ({\it i.e.}, for each vertex pair $(v_1,v_2)$ there are probability $p$ that there is an edge connecting these vertices) with $p=0.5$.

\subsubsection{Genuine multi-entropy}
\begin{figure}[t]
\centering
\includegraphics[width=.8\linewidth]{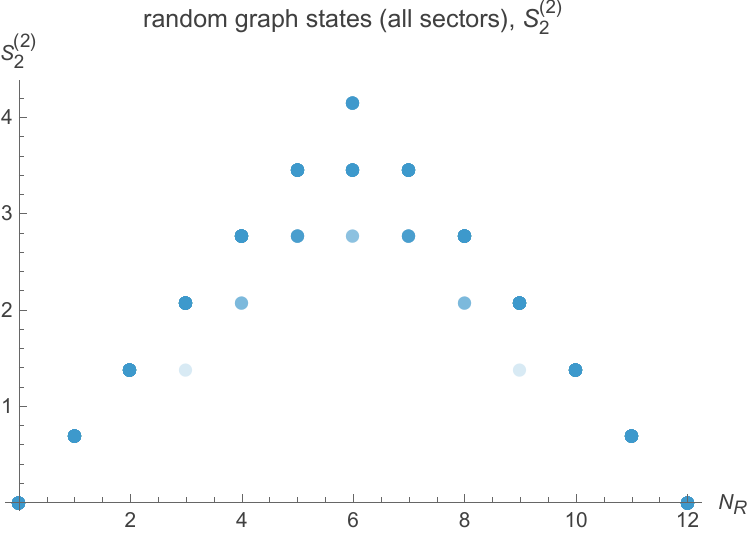}
\caption{Numerical evaluations of the second R\'enyi entropy (purity) of a random graph state on $N=12$ qubits. The detailed setup is the same as in Fig.~\ref{fig:graph_S34}.}
\label{fig:graph_S2}
\end{figure}

As a simple warmup, we compute the second R\'enyi entropy for random graph states in Fig.~\ref{fig:graph_S2}. 
We bipartition the vertex set $V$ into $R$ and $R^{c}$ with $|R|=N_R$ and $|R^{c}|=N-N_R$.
We find that R\'enyi entropy only takes values in integer multiples of $\log 2$, in agreement with \eqref{graphEE}.

We also see from Fig.~\ref{fig:graph_S2} that the entanglement entropy as a function of the $N_R$ behaves like the famous Page curve where it is a piecewise linear function that is proportional to $\min(N_R,N-N_R)$ in $\log 2$ unit. This should not come as a surprise since it is known that a large random Bernoulli matrix is almost always full rank\footnote{More precisely,  $\mathbf{E}({\rm Rank}_{\mathbb{Z}_2}(\Gamma))=\min(m,n)-O(1)$ for a $m\times n$ random Bernoulli matrix $\Gamma$ with $p=1/2$ and $m,n\gg 1$. For a proof of this fact see e.g. Ch.~3 of Ref.~\cite{Kolchin_1998}.}. Therefore, we conclude that random graph states have similar bipartite entanglement structure as uniform Haar random states for large $N$. As we will see soon below, this is not the case for higher partite entanglements.

Next, We compute the $\mathtt{k}=3,4$ (genuine) multi-entropies for random graph states. We find that, similar to the bipartite case, the multi-entropies also take discrete values, albeit no longer integer multiples of $\log 2$. The results are shown in Fig.~\ref{fig:graph_S34}.

Curiously, we noted that $\mathtt{k}=3$ genuine multi-entropy $\GM[3]_2$ is zero for all random graph states.
This is related to the fact that stabilizer states contain only GHZ-type genuine tripartite entanglement \cite{Bravyi:2005ztn,Nezami:2016zni} \footnote{Due to this property, it was recently argued that stabilizer states are not good models for holographic states in the sense that they have incorrect tripartite entanglement structure \cite{Balasubramanian:2025hxg,Akella:2025owv}.}.
More specifically, any tripartite stabilizer state $\ket{\psi}$ can be transformed into a collection of GHZ states $\ket{\rm GHZ}$, Bell pairs $\ket{\phi}$ and untangled qubits $\ket{0}$
\begin{equation}
\label{stab_decomp}
    \ket{\psi} \sim \ket{\rm GHZ}^{\otimes p} \otimes \ket{\rm \phi}_{AB}^{\otimes m_{AB}} \otimes \ket{\rm \phi}_{BC}^{\otimes m_{BC}} \otimes \ket{\rm \phi}_{AC}^{\otimes m_{AC}} \otimes \ket{0}^{\otimes s}
\end{equation}
under the action of local unitaries on individual subsystems.
For tripartite GHZ states,
$\GM[3]_n$ behaves as \cite{Liu:2024ulq}
\begin{equation}
  \label{GM3GHZ}
  \GM[3]_n(\ket{\text{GHZ}}) = \left(\frac{1+n}{n}-\frac{3}{2}\right)\log 2,
\end{equation}
which is zero for $n=2$. Hence the tripartite entanglement for graph states is not visible to $\GM[3]_2$ \cite{Iizuka:2025ioc}.

\begin{figure}[t]
\centering
\includegraphics[width=.8\linewidth]{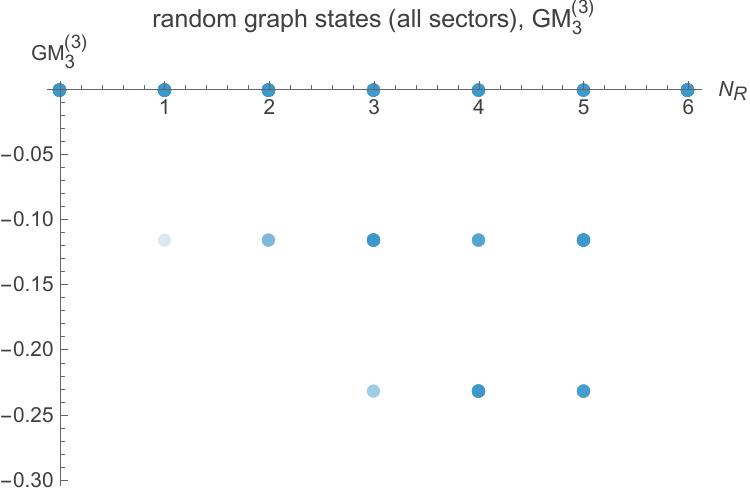}
\caption{Numerical evaluations of the third R\'enyi (genuine) multi-entropy $\GM[3]_3$   on $N=12$ spins. The detailed setup is the same as in Fig.~\ref{fig:graph_S34}.}
\label{fig:graph_GM33}
\end{figure}

\begin{figure*}[t]
\centering
\includegraphics[width=.4\linewidth]{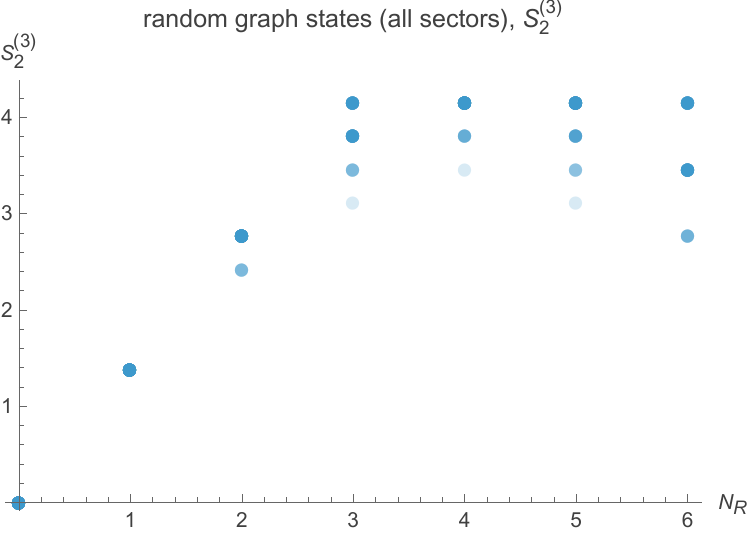}
\hspace{5mm}
\includegraphics[width=.4\linewidth]{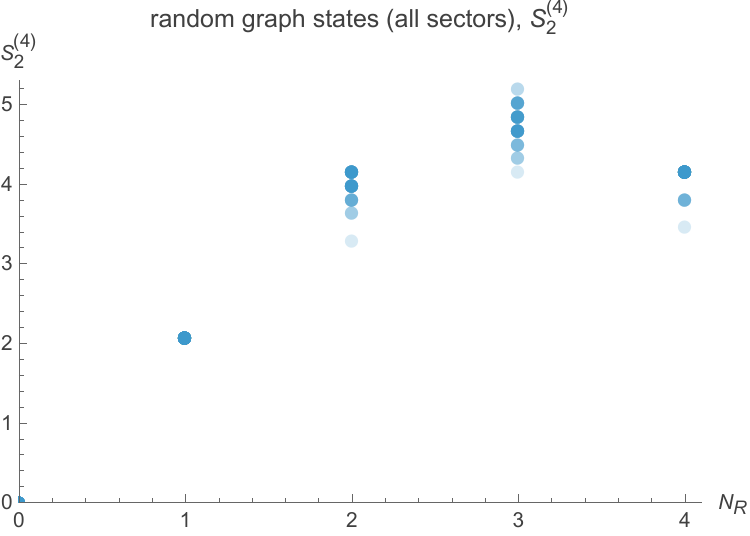}
\includegraphics[width=.4\linewidth]{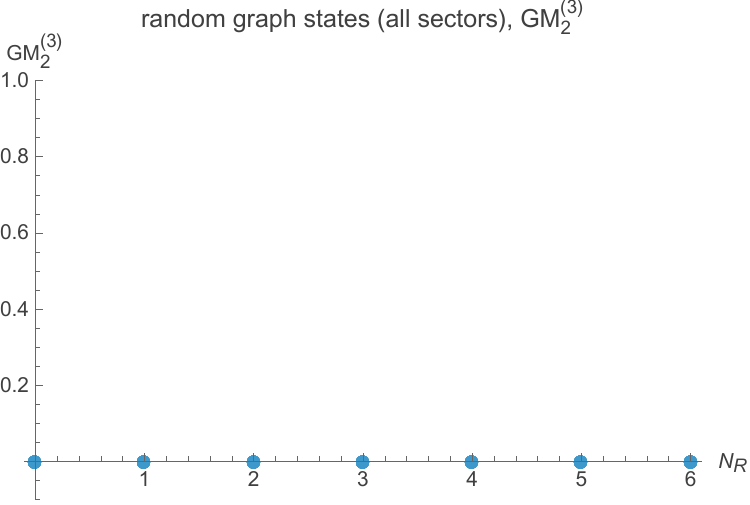}
\hspace{5mm}
\includegraphics[width=.4\linewidth]{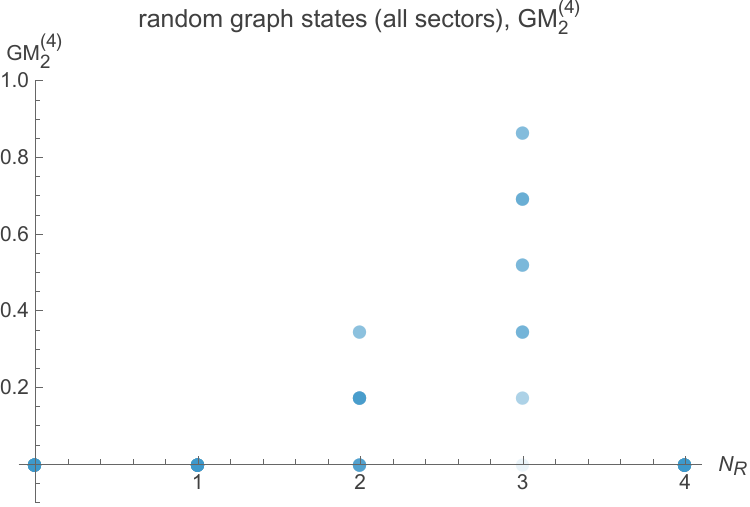}
\caption{Numerical evaluations of the $n=2$ multi-entropy (top row) and GM (bottom) of a random graph state on a spin chain with $N=12$ spins. For each $N_R$ we draw 50 samples from the Bernoulli graph distribution with $p=0.5$. The darkness of the data points represents the number of samples.}
\label{fig:graph_S34}
\end{figure*}

 To see nontrivial results we need to go to $n=3$. The number of GHZ states that can be distilled out of $\ket{G}$ is given by \cite{Bravyi:2005ztn}
\begin{equation}
  p=\log_2\left(\frac{|\mathcal{S}|}{|\Braket{\mathcal{S}_{AB},\mathcal{S}_{BC},S_{AC}}|}\right).
\end{equation}
Note that $|\Braket{\mathcal{S}_{AB},\mathcal{S}_{BC},S_{AC}}|\ne|S_{AB}||S_{BC}||S_{AC}|$ as opposed to the bipartite case.
Again, we can express this as the rank of the adjacency matrices of $G$. Let
\begin{equation}
  \Gamma =
  \begin{pmatrix}
    \Gamma_A & \Gamma^T_{AB} & \Gamma^T_{AC} \\
    \Gamma_{AB} & \Gamma_B & \Gamma^T_{BC} \\
    \Gamma_{AC} & \Gamma_{BC} & \Gamma_C
  \end{pmatrix}
\end{equation}
be the adjacency matrix of $G$. Then one has
\begin{align}
\label{gammarank}
     p &= \text{rank}_{\mathbb{Z}_2}(\Gamma_{A:BC})+\text{rank}_{\mathbb{Z}_2}(\Gamma_{B:AC})+\text{rank}_{\mathbb{Z}_2}(\Gamma_{C:AB}) \nonumber \\
  &\quad-\text{rank}_{\mathbb{Z}_2}(\tilde{\Gamma}),
\end{align}
\footnote{To the best of our knowledge, this expression has not been derived previously in the literature. We present a derivation of \eqref{gammarank}, as well as \eqref{graphEE} in Appendix~\ref{app:adjacency}. As a sanity check, we verified \eqref{GM3graph} numerically on graph states with various random graph ensembles.}where $\Gamma_{A:BC}=\begin{pmatrix}\Gamma_{AB}^T&\Gamma_{AC}^T\end{pmatrix}$ is the adjacency matrix between $A$ and $BC$ (likewise for $\Gamma_{B:AC}$ and $\Gamma_{C:AB}$), and $\tilde{\Gamma}$ is the diagonal-subtracted version of $\Gamma$, {\it i.e.}, 
\begin{equation}
  \tilde{\Gamma} =
  \begin{pmatrix}
    0 & \Gamma^T_{AB} & \Gamma^T_{AC} \\
    \Gamma_{AB} & 0 & \Gamma^T_{BC} \\
    \Gamma_{AC} & \Gamma_{BC} & 0
  \end{pmatrix}.
\end{equation}
Combining this with \eqref{GM3GHZ} we conclude that
\begin{equation}
\label{GM3graph}
  \GM[3]_3(\ket{G}) = -\frac{p}{6}\log 2,
\end{equation}
which is always negative and comes in integer multiples of $-\frac{1}{6}\log 2$. Indeed this is what we found numerically, as shown in Fig.~\ref{fig:graph_GM33}.

For $\mathtt{k=4}$, we find that the genuine multi-entropy $\GM[4]_2$ is positive in general. However, we noted that all the graph states we examined\footnote{For a sample size of $\sim10^4$ drawn from Bernoulli graph distribution with $p=0.5$ up to $N=12$.} satisfy the following property:
\begin{equation}
\label{GM4I3}
    4\GM[4]_2(\ket{G})+I_{3,n=2}(\ket{G})=0,
\end{equation}
which suggests that $\GM[4]_2$ for graph states satisfies
\begin{equation}
\label{GM4graph}
    4\GM[4]_2(\ket{G}) = -I_{3,n=2}(\ket{G}) =\frac{s}{4}\log 2,
\end{equation}
where
\begin{align}
    s &= \text{rank}_{\mathbb{Z}_2}(\Gamma_{AB:CD})+\text{rank}_{\mathbb{Z}_2}(\Gamma_{AC:BD})+\text{rank}_{\mathbb{Z}_2}(\Gamma_{AD:BC})\nonumber\\
    &\quad-\text{rank}_{\mathbb{Z}_2}(\Gamma_{A:BCD})-\text{rank}_{\mathbb{Z}_2}(\Gamma_{B:ACD})\nonumber\\
    &\quad-\text{rank}_{\mathbb{Z}_2}(\Gamma_{C:ABD})-\text{rank}_{\mathbb{Z}_2}(\Gamma_{D:ABC}) 
\end{align}

Eq.~\eqref{GM4I3} and \eqref{GM4graph} imply that the quadripartite entanglement of $\ket{G}$ is highly constrained. For example, the quadripartite W-state $\ket{W_4}=\frac{1}{2}\left( \ket{\uparrow\uparrow\uparrow\downarrow}+ \ket{\uparrow\uparrow\downarrow\uparrow}+ \ket{\uparrow\downarrow\uparrow\uparrow} + \ket{\downarrow\uparrow\uparrow\uparrow}\right)$ does not satisfy \eqref{GM4I3}. This is because W-state is not a graph state.
It can not be of GHZ type either, since the quadripartite GHZ state $\ket{\text{GHZ}_4}=(\ket{\uparrow\uparrow\uparrow\uparrow}+\ket{\downarrow\downarrow\downarrow\downarrow})/\sqrt{2}$ has negative $\GM[4]_2$ \cite{Iizuka:2025caq}.
In fact, as opposed to the tripartite case, it has been shown that there does not exist a decomposition for stabilizer states in terms of a finite list of elementary ``basis states'' like in \eqref{stab_decomp} for $\mathtt{k}\ge 4$ \cite{Englbrecht:2022fcc}. However our finding seems to indicate that these basis states, albeit an infinite list, all have  \emph{a very specific entanglement structure} given by \eqref{GM4I3} and \eqref{GM4graph}. This may be used to further classify the possible decompositions of higher-partite stabilizer states.

\subsubsection{Symmetry-resolved GM}
\begin{figure*}[h]
\centering
\includegraphics[width=.43\linewidth]{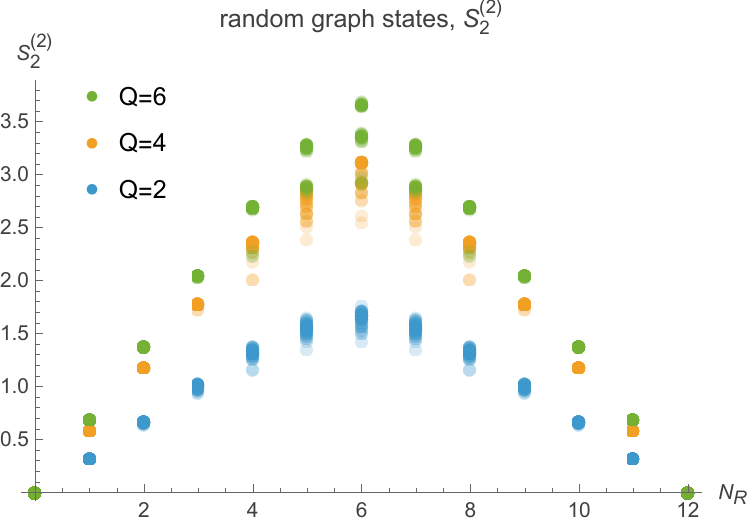}\\
\vspace{5mm}
\includegraphics[width=.4\linewidth]{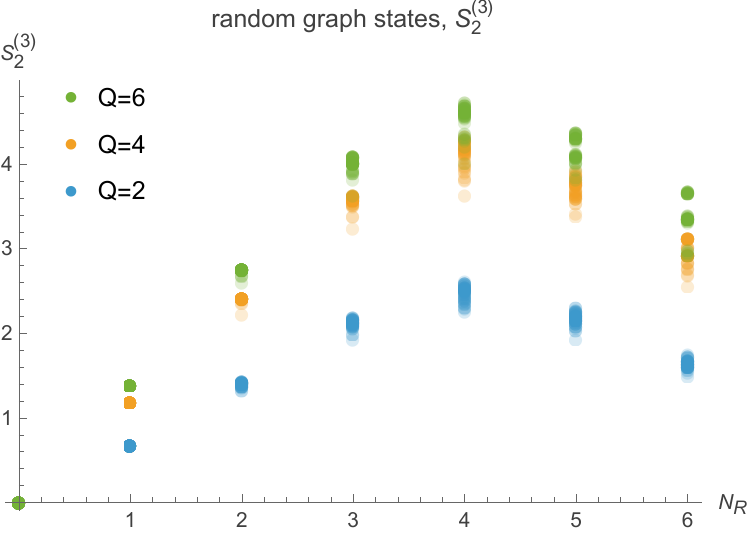}
\hspace{5mm}
\includegraphics[width=.4\linewidth]{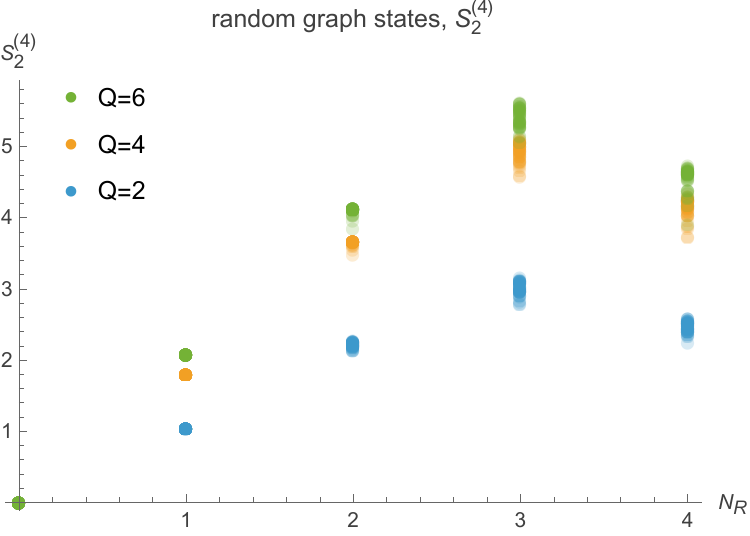}
\includegraphics[width=.4\linewidth]{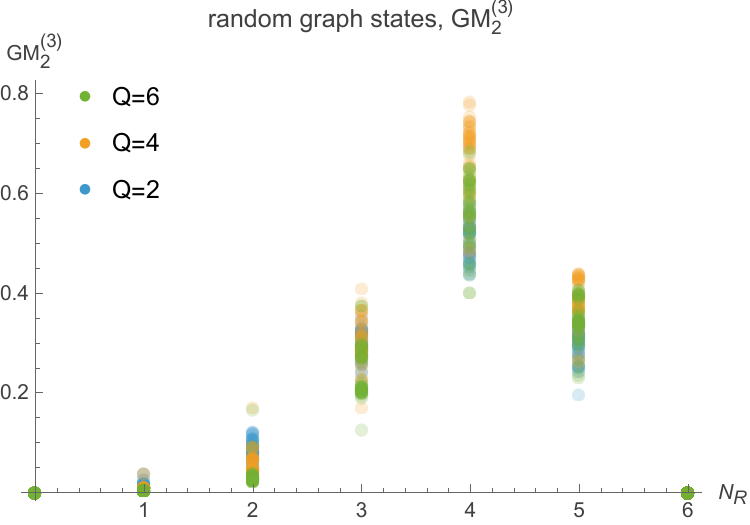}
\hspace{5mm}
\includegraphics[width=.4\linewidth]{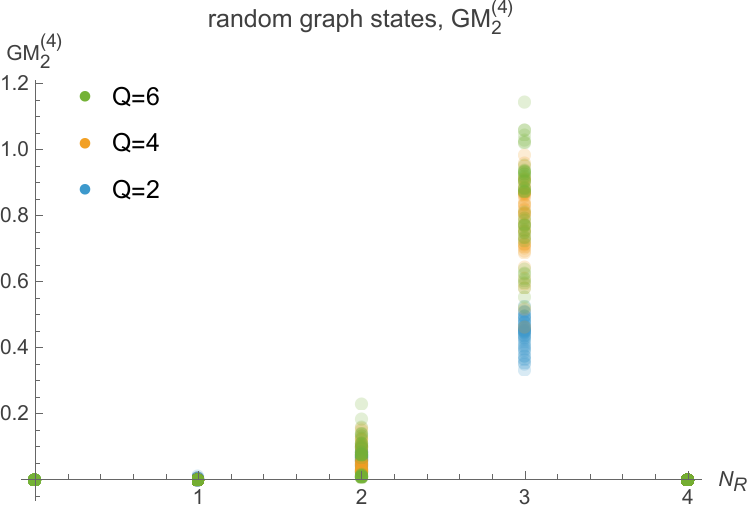}
\caption{Numerical evaluations of the symmetry-resolved second R\'enyi (genuine) multi-entropy of a random graph state on a spin chain with $N=12$ spins.
  For each graph state we obtained from the random graph $G$ we obtain the symmetry-resolved entropies by simply projecting down to fixed charged sectors $Q=2,4,6$.}
\label{fig:graph_SR}
\end{figure*}

We now look at symmetry-resolved entropies of random graph states.
In general, the state $\ket{G}$ for some random graph $G$ does not have a definite $U(1)$ charge. To overcome this issue we project $\ket{G}$ onto a charged sector by the corresponding projector $\Pi_q$. The results are shown in Fig.~\ref{fig:graph_SR}. 
Surprisingly, after the projection the discreteness of symmetry-resolved entropy ``smears out'' with characteristics more similar to true Haar random states.

The tripartite genuine multi-entropy $\GM[3]_2$ is no longer zero after the projection and features a similar shape akin to the Haar random state.
However, the detailed dependence on the global charge is different from true Haar random states. For example, the value of $\GM[3]_2$ for the $Q=4$ sector is larger than the $Q=6$ sector. The quadripartite genuine multi-entropy shows a similar behavior, where the shape of the curve for each charged sector is similar to Haar random states but the relative weight between individual sectors are different.

\section{Discussion}
\label{sec:summary}
In this letter, we examined the multi-partite entanglement structure of Haar random states and random graph states for a qubit system under a global $U(1)$ symmetry 
using multi-entropy and genuine multi-entropy. 
Note that throughout this work, Haar-random states are used as a convenient proxy for highly chaotic quantum systems. Physically, this should be understood as assuming approximate unitary design behavior up to the moment order required by the observable, rather than exact Haar randomness. 
The main advantage of these measures over entanglement entropy ($\mathtt{k}=2$) is that
entanglement entropy is sensitive only to the bipartite entanglement spectrum (the eigenvalues of the reduced density matrix) within each symmetry sector and therefore cannot determine genuine higher-partite structure. By contrast, the (genuine) multi-entropy with $\mathtt{k}\ge 3$ captures irreducible multi-partite correlations invisible to $\mathtt{k}=2$.
We derived analytical formulae for Haar random states in the thermodynamic limit and performed numerical simulations for both Haar random and graph states on a chain of $N=12$ qubits. For Haar random states, we find that the symmetry-resolved (genuine) multi-entropy curves closely resemble those of the full Haar random states without a global symmetry.

For graph states (without global symmetry), we verify the common lore that their higher-partite entanglement structure is both qualitatively and quantitatively different from that of a Haar random state. 
In particular, large $N$ graph states can look Haar-typical under entanglement entropy
for the bipartite case without the need to project. On the other hand, for multi-partite cases, 
they do not show Haar-typical behavior before the projection.
After projecting down to different symmetry sectors, we seem to recover more typicality in the sense that the higher-partite entanglement profile is more similar to a Haar random one, although there are still qualitative differences (for example the discreteness is not completely removed in Fig.~\ref{fig:graph_SR}).

Several open questions remain. First of all, it would be nice to extend the analytical analysis of the Haar random state to incorporate finite $N$ corrections. This is more difficult than the bipartite case because the expressions for multi-entropy as do not factorize into different local charge subsectors as opposed to the bipartite R\'enyi entropies. Another interesting parameter range is where $N\to\infty$ while keeping $N_R$ finite. This setup is interesting since it closely resembles the evaporation of an extremal charged black hole, where it is expected that the Hawking quanta should be non-typical \cite{Preskill:1991tb,Maldacena:1998uz,Page:2000dk,Brown:2024ajk}. 

For graph states, we were able to give explicit formulae for the (non-symmetry-resolved) multi-entropy in terms of their adjacency matrices for $\mathtt{k}=2,3$, and provided a guess for $\mathtt{k}=4$, {\it i.e.},  \ref{GM4graph}. A natural question is whether there exist similar formulae for higher $\mathtt{k}$. One must be cautious here, since based on \eqref{graphEE} and \eqref{GM3graph}, naively one could have guessed the following expression 
\begin{equation}
\label{graphGM4}
    \GM[4]_2(\ket{G}) \overset{?}{=} \frac{|\mathcal{S}|}{|\braket{\mathcal{S}_{ABC},\mathcal{S}_{ACD},\mathcal{S}_{ABD},\mathcal{S}_{BCD}}|}
\end{equation}
is true for $\mathtt{k}=4$. This particular form has been suggested as a measure for genuine multi-partite entanglement for stabilizer states \cite{Fattal:2004frh}. However, one can check that this is not the case. Instead \eqref{GM4graph} is  equal to
\begin{equation}
\label{graphGM4alt}
    \GM[4]_2(\ket{G}) = \frac{|\mathcal{S}||\mathcal{S}_{AB}||\mathcal{S}_{AC}||\mathcal{S}_{AD}|}{|\mathcal{S}_{ABC}||\mathcal{S}_{ABD}||\mathcal{S}_{ACD}||\mathcal{S}_{BCD}|}
\end{equation}
Thus there does not seem to be a clear pattern to proceed.

We should also note that our study of symmetry-resolved multi-entropy on random graph states has been limited to numerical simulations. It would be nice to perform an analytical analysis for this case. The projection down to individual each subsectors is not a stabilizer action, so a different approach is likely needed.

Finally, we discuss the possible lessons for black holes and holography. Our symmetry-resolved results indicate that imposing only $\order{1}$ charges (`a few hairs of black holes') leaves Haar-typical structure essentially intact even for multi-partite entanglement structure. Hence behavior that relies on typicality should be robust to a small set of conserved quantities during evaporation. By contrast, as we have seen in Eq.~\eqref{GM4I3} and \eqref{GM4graph}, code-like structure (stabilizer/graph families), even after projected, can significantly bias \emph{genuine} multi-partite entanglement.  
Toy models based purely on such code states may accurately capture error-correction features yet misrepresent higher-partite patterns that control the dynamics of black hole evaporation (e.g., the $\mathtt{k}$-dependent kink/onset scales and the sector dependence at fixed $Q$). Taken together, these observations make one point clear: a faithful model of black-hole evaporation requires a sharper, quantitative understanding of genuinely higher-partite ($\mathtt{k}>2$) entanglement, beyond what symmetry-resolved entanglement entropy alone can capture. 

\begin{acknowledgments}
  We thank Abhijit Gadde for the suggestion on the study of graph states.
  The work of N.I. was supported in part by MEXT KAKENHI Grant-in-Aid for Transformative Research Areas A “Extreme Universe” No. 21H05184 and by NSTC of Taiwan Grant Number 114-2112-M-007-025-MY3. 
  S.L. would like to thank the hospitality of NCTS and NTHU during the `NTU-NCTS Holography and Quantum Information Workshop', where part of this work is completed.
\end{acknowledgments}

\onecolumngrid
\appendix

\section{The uniform measure on a charged sector}
\label{app:measure}
We now demonstrate how \eqref{Qmeasure} is derived.
Any state $\ket{\psi}\in H(Q)$ can be decomposed into $\ket{\psi}=\sum_q\sqrt{p_q} \ket{\phi_q}$ with $\sum_qp_q=1$ and $\ket{\phi_q}\in H(q)$ is a normalized unit vector in the $q$-th subsector. 

Let's focus on a particular subspace $H(q)$ labeled by $q$.
Let $\ket{1},\ket{2},\cdots,\ket{d_q}$ be a basis for $H(q)$ where $d_q$ is the dimension of the $q$-th subsector. We can express the projection of $\ket{\psi}$ into $H(q)$ using
\begin{align}
    \Pi_q\ket{\psi} &= \psi_1\ket{1}+\psi_2\ket{2}+\cdots+\psi_{d_q}\ket{d_q} 
    = \sqrt{|\psi_1|^2+\psi_2|^2+\cdots+\psi_q|^2}\ket{\phi_q},
\end{align}
from which we can identify $p_q=\sum_{i=1}^{d_q}|\psi_i|^2$.
Likewise we can write the uniform measure on $H(q)$ as follows \footnote{To see that this is indeed the correct form of the measure, note that $\delta(|\phi_q|-1) d\phi_{1}d\bar{\phi}_{1} \cdots d\phi_{d_q}d\bar{\phi}_{d_q}$ is proportional to the uniform measure on the unit sphere of $S^{2d}$. The $p_q^{d_q-1}$  factor comes from the area of the $2d$ sphere with radius $\sqrt{p_q}$.}:
\begin{equation}
d\psi_1d\bar{\psi}_1\cdots d\bar{\psi}_{d_q} \propto p_q^{d_q-1} dp_q \delta(|\phi_q|-1) d\phi_{1}d\bar{\phi}_{1} \cdots d\phi_{d_q}d\bar{\phi}_{d_q} = p_q^{d_q-1}dp_q d\mu(\phi_q),
\end{equation}
where $\phi_i$ is the $i$-th component of $\ket{\phi_q}$ and $d\mu(\phi_q)$ is the uniform measure of $\phi_q$.
Repeating this for all subspaces $H(q)$ we obtain
\begin{equation}
  d\mu(\psi)  \propto  \left(\delta(\textstyle\sum_q p_q-1) \prod_q p_q^{d_q-1} dp_q\right) \prod_q d\mu(\phi_q)
\end{equation}
as claimed.

The normalization factor $\mathcal{Z}$ is fixed by requiring $\int d\nu=1$  and can be calculated to be \cite{Bianchi:2019stn}
\begin{equation}
   \mathcal{Z} = \frac{\prod_q \Gamma\big(d_q\big)}{\Gamma\big(\textstyle\sum_q d_q\big)}.
\end{equation}

\section{Entanglement of a graph state in terms of its adjacency matrix}
\label{app:adjacency}
In this appendix we demonstrate how to express the entanglement and tripartite multi-entropy of a graph state $\ket{G}$ in terms of the adjacency matrix of $G$.
Our strategy will involve deriving a relation between the cardinality of the stabilizer subgroup $\mathcal{S}_R$, which acts as identity on $\bar{R}$, and the rank of its adjacency matrices. \eqref{graphEE} and \eqref{gammarank} then follows from a straightforward application to \eqref{stabilizerEE} and \eqref{GM3GHZ}, which are valid for all stabilizer states.

Let $G=(V,E)$ be a graph. Recall that the stabilizer group of the graph state $\ket{G}$ is generated by
\begin{equation}
    S_{v} = \sigma_x^{(v)}\prod_{u\in N(v)} \sigma_z^{(u)},
\end{equation}
where $v\in V$ and $N(v)$ is the set of neighboring vertices of $v$. Any member of the stabilizer group $s\in \mathcal{S}$ then can be expressed as
\begin{equation}
    s = (S_{v_1})^{x_1}(S_{v_2})^{x_2}\cdots(S_{v_N})^{x_N},
\end{equation}
where $v_1,\cdots,v_N$ label the vertices in $G$ and $N=|V|$. $x_v\in\{0,1\}$ are a list of $N$ binary numbers. The upshot is that given the list $\{x_i\}$, we can uniquely specify an element in $\mathcal{S}$.

Consider some subset $R\subseteq V$. We wish to find the elements of $\mathcal{S}_R$, {\it i.e.}, the set of group elements that acts as identity on $\bar{R}=V\backslash R$. It is clear that we must have $x_i=0$ for $i\in \bar{R}$. However this is not sufficient, as $S_j$ with $j\in R$ has non-trivial action on $i$ if $j$ is connected to $i$ by an edge, unless the set of all such $j$'s that connect to $i$ is even. This is equivalent to the statement that $i$ has an even number of neighbors $j$ with $x_j=1$, or
\begin{equation}
\label{adjmod}
    \sum_{j\in R} \Gamma_{ij} x_j = 0 \mod 2,
\end{equation}
where $\Gamma_{ij}$ is the adjacency matrix of $G$. Therefore, we find that the elements of $\mathcal{S}_R$ is in one to one correspondence to the $\mathbb{Z}_2$ null vectors of the adjacency matrix $\Gamma_{R\bar{R}}$ between $R$ and $\bar{R}$, {\it i.e.}, 
\begin{equation}
    |\mathcal{S}_R| = 2^{|\text{Null}_{\mathbb{Z}_2}(\Gamma_{R\bar{R}})|} = 2^{|R|-\text{Rank}_{\mathbb{Z}_2}(\Gamma_{R\bar{R}})}.
\end{equation}
Since $|\mathcal{S}|=2^{|R|+|\bar{R}|}$ we have
\begin{equation}
    S_n(R)=\frac{1}{2}\log\left(\frac{|\mathcal{S}|}{|\mathcal{S}_R||\mathcal{S}_{\bar{R}}|}\right) = \text{Rank}_{\mathbb{Z}_2}(\Gamma_{R\bar{R}})\log 2.
\end{equation}

For the tripartite case $V=A\cup B \cup C$, we need to evaluate the dimension of the finitely generated subgroup $\mathcal{S}_{ABC}\equiv\braket{\mathcal{S}_{AB},\mathcal{S}_{BC},\mathcal{S}_{AC}}$.
Naively one would expect that $|\mathcal{S}_{ABC}|= |\mathcal{S}_{AB}||\mathcal{S}_{BC}||\mathcal{S}_{AC}|$. However this is not true since we have double counted the elements simultaneously satisfying an analog of \eqref{adjmod} for all of $A,B,C$. Let
\begin{equation}
  \Gamma =
  \begin{pmatrix}
    \Gamma_A & \Gamma^T_{AB} & \Gamma^T_{AC} \\
    \Gamma_{AB} & \Gamma_B & \Gamma^T_{BC} \\
    \Gamma_{AC} & \Gamma_{BC} & \Gamma_C
  \end{pmatrix}
\end{equation}
be the full adjacency matrix of $G$. These elements correspond to the null vectors of the diagonal-subtracted adjacency matrix
\begin{equation}
  \tilde{\Gamma} \equiv
  \begin{pmatrix}
    0 & \Gamma^T_{AB} & \Gamma^T_{AC} \\
    \Gamma_{AB} & 0 & \Gamma^T_{BC} \\
    \Gamma_{AC} & \Gamma_{BC} & 0
  \end{pmatrix}
\end{equation}
Therefore,
\begin{align}
\begin{split}
    |\mathcal{S}_{ABC}| &= \frac{|\mathcal{S}_{AB}||\mathcal{S}_{BC}||\mathcal{S}_{AC}|}{2^{\text{Null}_{\mathbb{Z}_2}(\tilde{\Gamma})}} \\
    &=\frac{2^{\text{Null}_{\mathbb{Z}_2}(\Gamma_{A:BC})}2^{\text{Null}_{\mathbb{Z}_2}(\Gamma_{B:AC})}2^{\text{Null}_{\mathbb{Z}_2}(\Gamma_{C:AB})}}{2^{\text{Null}_{\mathbb{Z}_2}(\tilde{\Gamma})}} \\
    &=\frac{2^{2|V|-\text{Rank}_{\mathbb{Z}_2}(\Gamma_{A:BC})-\text{Rank}_{\mathbb{Z}_2}(\Gamma_{B:AC})-\text{Rank}_{\mathbb{Z}_2}(\Gamma_{C:AB})}}{2^{|V|-\text{Rank}_{\mathbb{Z}_2}(\tilde{\Gamma})}},
\end{split}
\end{align}
where $\Gamma_{A:BC}=\begin{pmatrix}\Gamma_{AB}^T&\Gamma_{AC}^T\end{pmatrix}$ is the adjacency matrix between $A$ and $BC$ and likewise for $\Gamma_{B:AC}$ and $\Gamma_{C:AB}$.
Thus
\begin{equation}
    \log_2\left(\frac{|\mathcal{S}|}{|\mathcal{S}_{ABC}|}\right) = \text{Rank}_{\mathbb{Z}_2}(\Gamma_{A:BC})+\text{Rank}_{\mathbb{Z}_2}(\Gamma_{B:AC})+\text{Rank}_{\mathbb{Z}_2}(\Gamma_{C:AB})-\text{Rank}_{\mathbb{Z}_2}(\tilde{\Gamma}).
\end{equation}

\twocolumngrid
\bibliography{Refs}
\bibliographystyle{JHEP}

\end{document}